\documentclass[iop]{emulateapj}

\usepackage{natbib}
\usepackage{bm}
\usepackage{amsmath}
\usepackage{rotating}
\usepackage{gensymb}
\shorttitle{MHD Wind-Cloud Interactions}
\shortauthors{Cottle et al.}

\begin{document}

\title{The Launching of Cold Clouds by Galaxy Outflows III: The Influence of Magnetic Fields}


\author{J'Neil Cottle\altaffilmark{1}, Evan Scannapieco\altaffilmark{1}, Marcus Br\"uggen\altaffilmark{2}, Wladimir Banda-Barrag\'an\altaffilmark{2}, and Christoph Federrath\altaffilmark{3}}

\altaffiltext{1}{School of Earth and Space Exploration, Arizona State University, Tempe AZ}
\altaffiltext{2}{Hamburger Sternwarte, Universit\"at Hamburg, Gojenbergsweg 112, D-21029, Hamburg, Germany}
\altaffiltext{3}{Research School of Astronomy and Astrophysics, Australian National University, Canberra, ACT 2611, Australia}

\begin{abstract}

Motivated by observations of outflowing galaxies, we investigate the combined impact of magnetic fields and radiative cooling on the evolution of cold clouds embedded in a hot wind. We perform a collection of three-dimensional adaptive mesh refinement, magnetohydrodynamical simulations that span two resolutions, and include fields that are aligned and transverse to the oncoming, super-Alfv\'enic material. Aligned fields have little impact on the overall lifetime of the clouds over the non-magnetized case, although they do increase the mixing between the wind and cloud material by a factor of $\approx 3.$ Transverse fields lead to magnetic draping, which isolates the clouds, but they also squeeze material in the direction perpendicular to the field lines, which leads to rapid mass loss. A resolution study suggests that the magnetized simulations have somewhat better convergence properties than non-magnetized simulations, and that a resolution of 64 zones per cloud radius is sufficient to accurately describe these interactions. We conclude that the combined effects of radiative cooling and magnetic fields are dependent on field orientation, but are unlikely to enhance cloud lifetimes beyond the effect of radiative cooling alone.

\end{abstract}


\section{Introduction}
Understanding the evolution and disruption of wind-swept clouds is essential to understanding the circumgalactic medium (CGM), as winds driven by star formation and supernovae accelerate dense clouds past the limits of the galactic plane. These outflowing winds have long been considered theoretically \citep[e.g.][]{Chevalier1985, MacLow1999, Murray2005, Scannapieco2010, Sur2016, Scannapieco2017}, and observations have provided evidence for both their multiphase nature \citep[e.g.][]{Veilleux2005, Sturm2011, Meiring2013, Bolatto2013, Kacprzak2014, Rubin2014} and their impact on galactic evolution and star formation \citep[e.g.][]{Tremonti2004,Oppenheimer2010,Dave2011,Lu2015,Agertz2015}. However, the details of the interaction between the winds and entrained clouds have been difficult to investigate without the use of numerical studies.

In the purely hydrodynamical regime, \citet{Klein1994} showed that such clouds are accelerated over time-scales $\approx 3-4$ times longer than the cloud-crushing time (hereafter, t${\rm cc}$), which is defined as the time taken by an internal shock to travel across one cloud radius. However, further studies \citep[e.g.][]{Poludnenko2002, Pittard2009, Fragile2005, BandaBarragan2019} indicated that shocks and dynamical instabilities quickly destroy the clouds on timescales that are too short for the clouds to reach the speeds and distances at which they are observed in galactic outflows.

Studies focusing on the influence of other effects, such as radiative cooling \citep[e.g.][hereafter Paper I]{Schiano1995, Cooper2009, Schneider2017, McCourt2018, Gronke2018, Sparre2019,  Scannapieco2015} and thermal conduction, have also been carried out \citep[e.g.][hereafter Paper II]{Orlando2005, Bruggen2016}. In the cooling case, cloud disruption is delayed by the suppression of shock heating, which is the dominant disruption mechanism in cases in which the exterior flow is supersonic. In the case of thermal conduction, cloud disruption is delayed by the presence of an evaporative layer, which compresses the cloud and protects it from shredding by the exterior flow. However, in both scenarios, the clouds eventually fragment into smaller cores \citep{McCourt2018, Sparre2019} or condense into filaments, and the cloud cross-sections are too small to be accelerated by ram pressure to the extent observed.

The influence of magnetic fields on the wind-cloud interaction introduces a mechanism to balance the acceleration and destruction of the clouds. In early two-dimensional magnetohydrodynamical (MHD) studies \citep[e.g.][]{MacLow1994, Jones1996, Miniati1999} it was shown that uniform magnetic fields transverse to the flow are likely able to create a magnetic `bumper' at the front of the cloud and potentially reduce the effect of the instabilities that destroy the cloud. On the other hand, in the case of fields aligned with the flow, the wind was found to have a similar disruptive effect on the cloud as in the hydrodynamic case.

Continued studies in, both, 2D \citep[][]{Orlando2008, Pittard2009, Pittard2010} and 3D \citep[][]{Gregori2000, Shin2008, Pittard2016, Gronnow2017}, have considered both aligned and transverse field orientations, as well as explored the impact of varying the wind Mach number \citep{vanLoo2007}, magnetic field strength \citep{McCourt2015} and turbulence \citep[][]{BandaBarragan2018, Li2020}. A few have also investigated the effect of oblique fields \citep[e.g.][]{BandaBarragan2016,Gronnow2018}. These studies have found that Kelvin-Helmholtz instabilities are reduced in the presence of strong magnetic fields. Aligned magnetic fields have the capability to form a high magnetic pressure flux rope, while transverse fields are stretched along the front of the cloud, resulting in a magnetic pressure that is comparable to the ram pressure from the wind. Self-contained and turbulent magnetic fields have been found to suppress the disruption of the clouds and result in smaller fragments comoving with the wind \citep{Li2013, McCourt2015, BandaBarragan2018}.

It is clear that magnetic fields play an important role in the evolution of the entrained clouds, though most studies have been limited to the early stages of the interaction. As clouds are accelerated through the wind, simulation domains have been too small to follow them for long enough to fully understand the evolution of the wind-cloud interaction. In addition, many MHD studies have focused on models without radiative cooling, such that the combined effects of radiative cooling and magnetic fields have not been well constrained. In the few studies that have considered both \citep[i.e.][]{Johansson2013, McCourt2015, Gronke2019}, the parameter space of cooling timescales  and field orientations has not been fully investigated. 

Here we consider both the effects simultaneously, making comparisons across two magnetic field orientations and highlighting the impact of orientation on cooling efficiency.  We consider the dependence of the cloud evolution on spatial resolution as well as the stability of MHD clouds as compared to the non-magnetized case. Radiative cooling is treated the same throughout all simulations. We track the clouds for several cloud crushing times with the use of a frame-changing routine in order to study the long-term evolution (Paper I, Paper II). 

The structure of this paper is as follows. In Section \ref{sec:sim} we discuss the simulations and the physics relevant to cloud evolution and the parameter space. In Section \ref{sec:results} we discuss the results of the simulations emphasizing on the effects of magnetic fields, radiative cooling, and numerical resolution. We conclude in Section \ref{sec:summary} with a discussion and summary.

\begin{table}
\caption{Absolute Values of Wind and Cloud Parameters \label{tab:wind} }
\centering
\begin{tabular}{lc}
Variable & Value\\

\hline
$R_{\rm cloud}$ & 100 pc\\
$\rho_{\rm cloud}$ & $10^{\-24}$ g cm$^{-3}$ \\
$\rho_{\rm wind}$ & $10^{\-27}$ g cm$^{-3}$ \\
$T_{\rm hot}$ & 10$^{7}$ K \\
$T_{\rm cloud}$ & 10$^{4}$ K \\
$v_{\rm hot}$ & 1700 km/s \\
$t_{\rm cc}$ & 1.8 Myr \\
$N_{\rm cool}$ & $10^{17.5}$ cm$^{-2}$\\
$t_{\rm cool}$ & 1.84 yr \\

\end{tabular}
\end{table}

\section{Simulations}\label{sec:sim}
We performed a suite of MHD simulations of wind-cloud interactions including radiative cooling, using the code FLASH \citep[version 4.0.1][]{Fryxell2000, Dubey2008}. These simulations were done in three dimensions and made use of the HLL3R Riemann scheme \citep{Waagan2011}, which provides a stable solution in problems that involve strongly magnetized flows and high Mach numbers, with improved efficiency over the standard solvers within FLASH. Divergence cleaning is implemented with the existing scheme within FLASH; a parabolic cleaning method \citep{Marder1987}.

The simulations assumed an initial cloud radius of 100 parsec, a cloud temperature of 10$^4$ K, a mass density of $\rho = 10^{-24}$ g cm$^{-3}$ and a mean particle mass of $\mu = 0.6$. Initially, the cloud was positioned at (0,0,0) within the domain covering $-800 \times 800$ parsec in $x$ and $z$ and $-666 \times 1333 $ parsec in $y$, which was the direction of the hot, outflowing material. The interaction at the $y$-boundary was defined by a condition where the incoming material is added to the grid and given the same values of density, velocity ($v_{\rm hot}$), sound speed ($c_{\rm s,hot}$) and magnetic pressure as the initial wind conditions. For all other boundaries the FLASH ``diode" condition was used, which assumes the gradient normal to the edge of the domain to be zero for all variables except pressure and only allows material to flow out of the grid.

In order to resolve instabilities along the boundary of the cloud without drastically increasing the computation time, the simulations make use of FLASH's adaptive mesh refinement (AMR) capabilities \citep{Berger1989}. 
As in our previous studies, cells were refined according to the magnitude of the second derivative of density and temperature of the gas, we but also adopted a set of additional refinement and de-refinement criteria, cho- sen to minimize the computational cost of the simulation while at the same time maintaining the most accurate results possible in the spatial regions that are the most important to the evolution of the cold cloud (see Paper 1 for details). In the high-resolution case, five levels of refinement are used, with the cloud gas maintaining the highest level through the simulation. In this case, the lowest level of refinement produces 4 cells per initial cloud radius while the highest level of refinement provides 64 cells per cloud radius. For the low-resolution simulations only four levels of refinement are used with 4 cells per cloud radius at the lowest level and 32 cells per cloud radius at the highest level.

In order to follow the disruption of the clouds over long timescales, it was necessary for the simulations to shift frames as the cloud is accelerated by the wind. This was implemented with the use of an automated frame-change routine originally discussed in Paper I. Similarly, we also use a scalar to track cloud material, $C_{\rm cloud}$. Initially, this scalar is set to 0 within the wind and 1 in the cloud. As the gases mix, the scalar reflects the fraction of material within each cell that originated within the cloud. 

\subsection{Physics of Cloud Evolution}
There are two key timescales relevant to the evolution of a cloud embedded within a magnetized hot wind. The first, the cloud-crushing time, effectively describes the amount of time it would take the initial internal shock to travel halfway through the cloud. It is given by

\begin{equation}
t_{\rm cc} = \frac{R_c \chi_0^{1/2}}{v_{\rm hot} },
\end{equation}
\\
which, for a consistent cloud radius, is dependent only on the velocity of the wind, $v_{\rm hot}$, and the density ratio between the cloud and the wind, $\chi_0$ \citep[e.g.][]{Klein1994}.

In addition, the cooling time, which determines the time for the cloud to radiate away its thermal energy is given by
\begin{equation}
t_{\rm cool} = \frac{(3/2)n_c k_B T}{\Lambda(T) n_{e,c} n_{i,c}},
\end{equation}
\\
where $T$ is the temperature and $\Lambda(T)$ is the equilibrium cooling function at $T$ with $n_{c}$, $n_{e,c}$ and $n_{i,c}$ being the total, electron and ion number densities within the cloud. The cooling rate is taken from the tables constructed by \citet{Wiersma2009} with the assumption that the material is always solar metallicity. If the ratio of $t_{\rm cool}/t_{\rm cc} = N_{\rm cool}/(n_{i,c} r_c)$ with $N_{\rm cool} \equiv 3 k_B T v_{\rm hot} n_c (2 \Lambda \chi^{1/2} n_{e,c})^{-1}$ is below one, then cooling will have a significant influence on the evolution of the cloud as it will have a chance to cool prior to being disrupted by the shock. For these simulations $t_{\rm cool} / t_{\rm cc} \approx 1 \times 10^{-6} $ implying very efficient cooling through the evolution of the clouds.

While the absolute value of the cloud crushing time changes with the radius of the cloud, the ratio $t_{\rm cool}/t_{\rm cc}$ and therefore the evolution of the cloud, is only dependent on $N_{\rm cool}$ and $M_{\rm hot}$. When considered in units of the cloud crushing time, the evolution of the cloud is not dependent on the size of the cloud for a given $N_{\rm cool}$ and $M_{\rm hot}$. A smaller or denser cloud will evolve in a longer amount of absolute time but will reflect the same evolution in units of the cloud crushing timescale as a larger cloud with the same $N_{\rm cool}$ and $M_{\rm hot}$.  Absolute values for these particular clouds are listed in Table \ref{tab:wind}.

An important relation for magnetic fields is the plasma $\beta$, the ratio of the thermal and magnetic pressures
\begin{equation}
\beta = \frac{P_{\rm th}}{P_{\rm mag}} = \frac{(\rho/ \mu m_p) k_B T}{B^2/(2\mu_0)},
\end{equation}
with $\rho$ the density, $\mu$ again the mean particle mass, $T$ the temperature, $B$ the magnetic field strength and the constants being proton mass ($m_p$), the Boltzmann constant ($k_B$) and magnetic permeability of free space ($\mu_0$). This is one of the parameters used to describe the magnetic fields within the simulations. 

The ideal system of MHD equations with radiative cooling solved by FLASH in conservation form, with $I_3$ denoting the $3 \times 3$ identity matrix, is,
\begin{equation}
\rho_t + \nabla \cdot (\rho \bm{u}) = 0,
\end{equation}
\begin{equation}
(\rho \bm{u})_t + \nabla \cdot \left[ \rho \bm{u} \times \bm{u} + \left(p + \frac{1}{2} |\bm{B}|^2 \right) I_3 - \bm{B} \times \bm{B} \right] = 0,
\end{equation}
\begin{equation}
E_t + \nabla \cdot  \left[ \left(E + p + \frac{1}{2} |\bm{B}|^2 \right)\bm{u} - \left(\bm{B} \cdot \bm{u} \right)\bm{B} \right] + \dot{E}_{\rm cool} = 0,
\end{equation}
\begin{equation}
\bm{B}_t + \nabla \cdot (\bm{B} \times \bm{u} - \bm{u} \times \bm{B}) = 0,
\end{equation}
\begin{equation}
\nabla \cdot \bm{B} = 0,
\end{equation}
with $\rho$ the density, $\bm{u}$ the velocity, $p = k_B T \rho/(\mu m_p)$ the pressure and $E = p/(\gamma - 1) + \frac{1}{2} \rho |\bm{u}|^2 + \frac{1}{2} |\bm{B}|^2$ the total energy density. The solver presented in \citet{Waagan2011} makes use of a second-order scheme with an entropy-stable approximate Riemann solver. This solver uses primitive variables with relaxation solvers which helps reproduce material contact discontinuities. It has been found to have increased efficiency and stability especially for high Mach number flows and low plasma $\beta$. Both of these are directly applicable to this study.

Our simulations also account for radiative cooling. In the optically-thin limit, the additional change in energy due to cooling, the radiated energy per unit mass $\dot{E}_{\rm cool}$, is given by
\begin{equation}
\dot{E}_{\rm cool} = (1-Y)\left(1 - \frac{Y}{2}\right) \frac{\rho \Lambda}{(\mu m_p)^2},
\end{equation}
where $\rho$ is the density, $Y=0.24$ is the helium mass fraction, $\mu=0.6$ is the mean atomic mass, $m_p$ is the proton mass and $\Lambda$ is the cooling rate as a function of temperature and metallicity. 
Heating by a photoionizing background was not included in the calculations, and sub-cycling was implemented \citep{Gray2010} along with a cooling floor at $T=10^4$K.

\subsection{Parameters}

The wind is described by three parameters, $\mathcal{M}_{\rm hot}$, $v_{\rm hot}$ and $T_{\rm hot}$. The Mach number of the inflowing material, $\mathcal{M}_{\rm hot}$, reflects the conditions at a particular radius from the outflowing region while the velocity of this hot medium, $v_{\rm hot}$, captures both the energy and mass input from the wind \citep{Chevalier1985}. The temperature of the wind is denoted by $T_{\rm hot}$ while the cloud is always at an initial temperature of $10^4$ K, the minimum temperature attainable with atomic cooling. For a cloud at this initial temperature, the Jeans length is $\lambda_J \approx 2$ kpc, much larger than the size of the clouds considered. This implies that the clouds must be confined by pressure in order to be in equilibrium at the start of the simulation and means that self-gravity (not included) is not important for this particular setup. Due to this, the ratio of the cloud density to the wind density, $\chi_0$, is equal to the ratio of the temperatures of the wind and cloud.

The magnetic fields are determined by two parameters, plasma beta, $\beta$, the ratio of thermal to magnetic pressure, and the angle with respect to the wind velocity. Since the wind and cloud are originally in pressure equilibrium, the initial plasma $\beta$ holds for all phases. We adopt $\beta = 10$; corresponding to a field strength of 1.86 $\mu$G for most runs with an additional two runs with an initial $\beta = 1$ (5.88 $\mu$G). These values reflect the lower limits of magnetic fields seen in observations of galactic outflows \citep{Adebahr2018}. 
The Alfvenic Mach number for these simulations is $\sim 91$ and $\sim 28$ for $\beta =10$ and $\beta = 1$ respectively. 

We then consider two different orientations for the field: aligned and transverse. The aligned case implies an angle of 0\degree \, between the field lines and the wind velocity with the only component of the field being in the $y$-direction. The transverse case describes field lines perpendicular to the wind velocity with the only component of the field being in the $x$-direction. Initially the $z$-component of the field is always taken to be zero.

A table of parameters is shown in Table \ref{tab:params} outlining the name of the simulation, magnetic field direction, resolution and the inclusion of radiative cooling. We focus on the primary case with a Mach number of 3.5 with wind parameters of $T_{\rm hot} = 10^7$ K, $v_{\rm hot} = 1700$ km/s, and $\chi_{0} = 1000$. We also consider the complementary, non-MHD run discussed in Paper I. The speed of the hot phase of the Milky Way's wind has been estimated to be upwards of 1000 km/s \citep{Carretti2013, McClure2013}, while these are the upper estimates, this study is relevant to the hot phase in galactic winds as well as applicable to the general study of the interaction of magnetized clouds and hot winds.

\begin{table}
\caption{Simulation parameters \label{tab:params} }
\centering
\begin{tabular}{lcccccccc}
Name & B-field & $\beta$ & Resolution & Cooling \\
& Angle & & (cells/r$_{\rm cloud}$) & \\
\hline

H-rad-lr & - & & 32 & \checkmark \\
H-rad-hr & - & & 64 & \checkmark\\
H-nonrad-hr & - & & 64 & \\
A-rad-lr & Aligned & 10 & 32 & \checkmark\\
A-nonrad-lr & Aligned & 10 & 32 & \\
A-rad-hr & Aligned &10 & 64 & \checkmark\\
A-nonrad-hr & Aligned & 10 & 64 & \\
A-B1-rad-hr & Aligned & 1 & 64 & \checkmark\\
T-rad-lr & Transverse & 10 & 32 & \checkmark\\
T-nonrad-lr & Transverse & 10 & 32 & \\
T-rad-hr & Transverse & 10 & 64 & \checkmark\\
T-nonrad-hr & Transverse & 10 & 64 & \\
T-B1-rad-hr & Transverse & 1 & 64 & \checkmark\\

\end{tabular}
\end{table}

\section{Results} \label{sec:results}

We carried out 10 simulations, which span the parameters in Table \ref{tab:params}.  These includes 8 MHD runs and 2 pure-hydro runs that use the standard directionally split Piecewise-Parabolic Method \citep{Colella1984} and complement the run carried out in Paper I.   In that paper, we showed that in the radiative non-MHD case  the evolution of the cloud converges at a resolution of typo $R_{\rm cloud}/64$. In order to test convergence, while keeping computational costs manageable, the MHD cases are run on a base grid of 64 $\times$ 80 $\times $ 64 with three and four additional levels of refinement for the low and high resolution runs, respectively. At the most refined level this corresponds to resolutions of $R_{\rm cloud}/32$ and $R_{\rm cloud}/64$. The domain extends over a physical volume of $-800$ to $800$ parsec in $x$ and $z$ and $ -666$ to $1333 $ parsec in $y$, the direction of the hot outflowing material.

\subsection{Impact of Radiative Cooling}
In the most basic wind-cloud scenario, a non-magnetized wind without cooling, the cloud is destroyed by the reflected shock that is produced as the initial shock wraps around the cloud, at about 2 $t_{\rm cc}$ \citep[e.g.][]{Klein1994}. This shock travels upstream, and works to tear apart the cloud, leading to catastrophic mass loss.

\begin{figure*}
\includegraphics[angle=0,width=\textwidth]{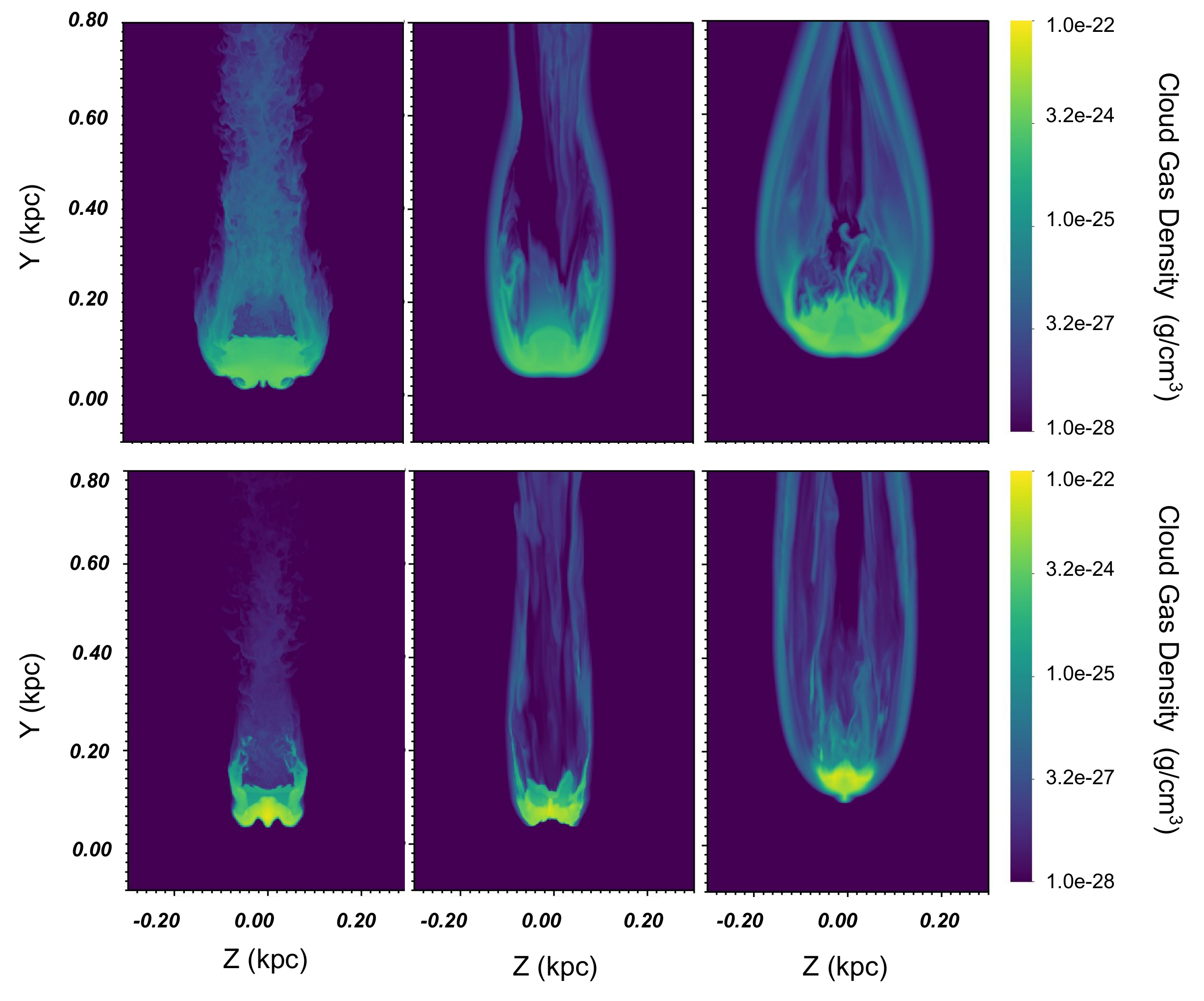}
\caption{Slices along the x-axis of the cloud density comparing the non radiative runs (top) with radiative runs (bottom) at 2 $t_{\rm cc}$. The hydrodynamic runs are shown on the left, aligned fields in the middle and transverse fields on the right.  All densities are given in g/cm$^{-3}$ and all lengths are given in kpc. These are zoomed in images of the more extended computational domains.
\label{fig:adiFigs1}}
\end{figure*}

\begin{figure*}
\includegraphics[angle=0,width=\textwidth]{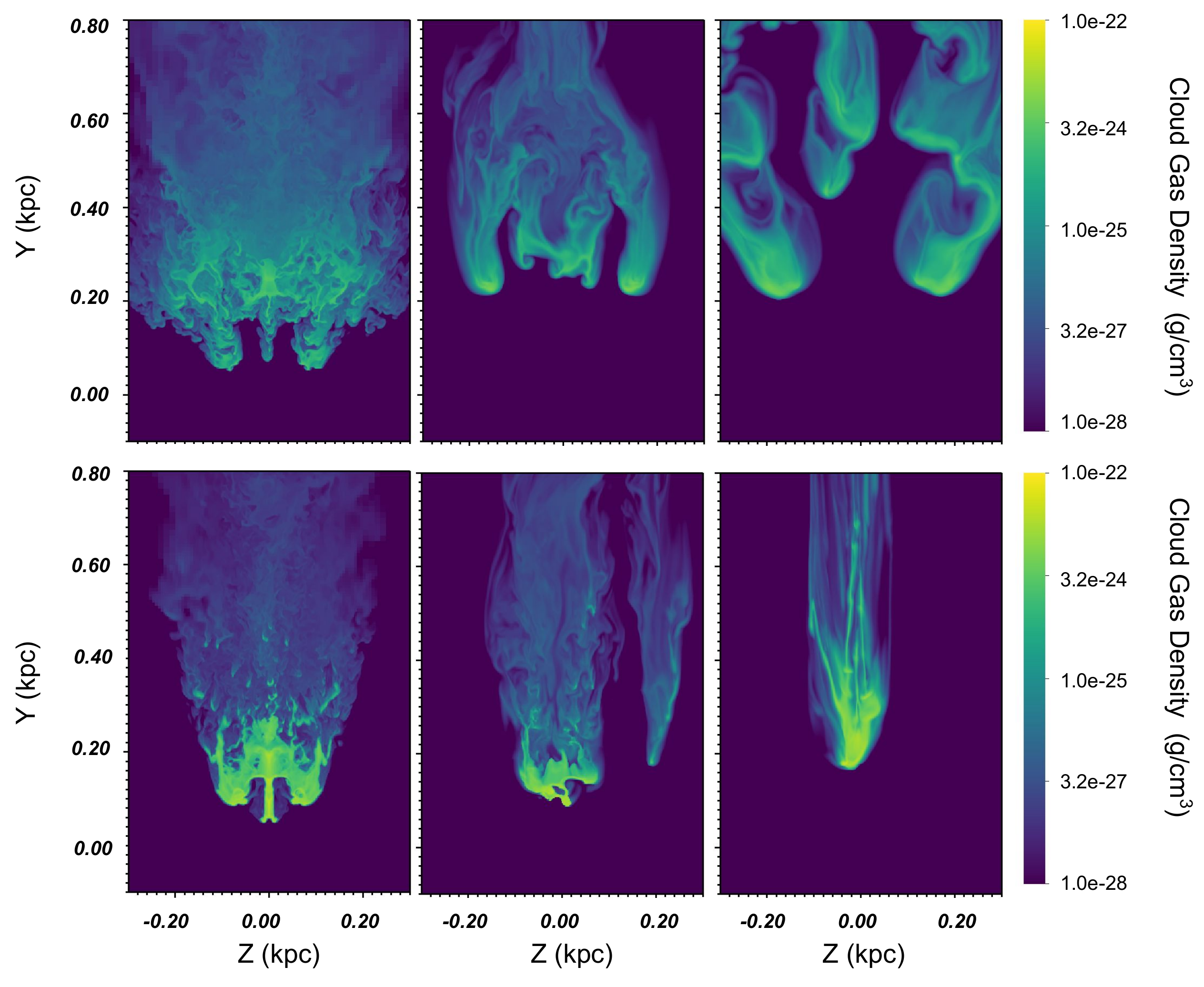}
\caption{Same as Figure \ref{fig:adiFigs1} but at 4 $t_{\rm cc}$.
\label{fig:adiFigs2}}
\end{figure*}

\begin{figure*}
\includegraphics[angle=0,width=\textwidth]{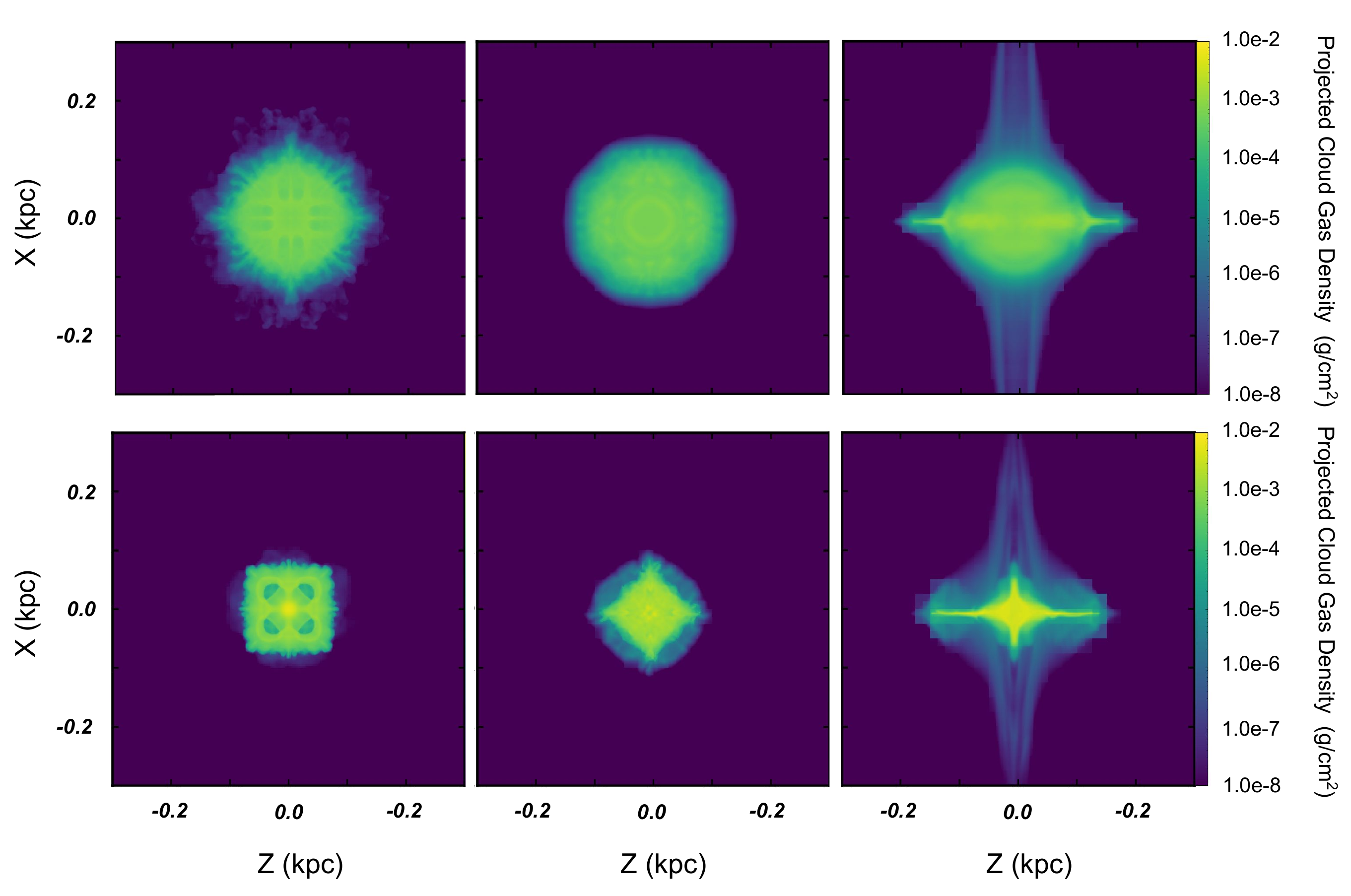}
\caption{Projections along the y axis of the cloud density comparing the non-radiative runs (top) with the radiative runs (bottom) at 2 $t_{\rm cc}$. The hydrodynamic runs are shown on the left, aligned fields in the middle and transverse fields on the right.  All column densities are given in g/cm$^{-2}$ and all lengths are given in kpc. These are zoomed-in images of the more extended computational domains.
\label{fig:projFigs1}}
\end{figure*}

\begin{figure*}
\includegraphics[angle=0,width=\textwidth]{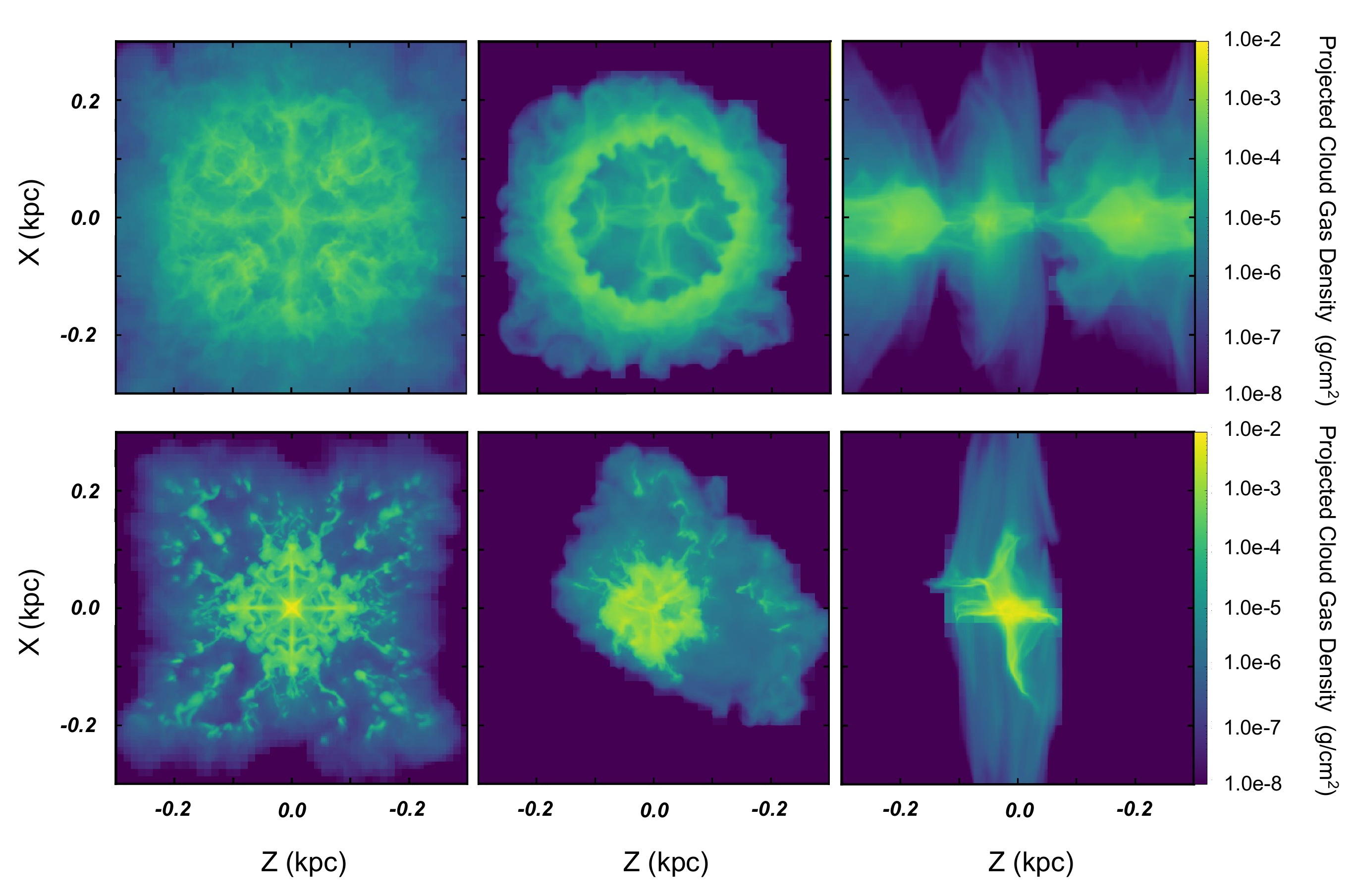}
\caption{Same as Figure \ref{fig:projFigs1} but at 4 $t_{\rm cc}$.
\label{fig:projFigs2}}
\end{figure*}

Shown in Figures \ref{fig:adiFigs1} and \ref{fig:adiFigs2}, are comparisons between the runs without cooling and radiative runs for the hydrodynamic and MHD simulations. Two times are shown, 2 $t_{\rm cc}$ (Figure \ref{fig:adiFigs1}) and 4 $t_{\rm cc}$ (Figure \ref{fig:adiFigs2}), with the runs without cooling on the top and the radiative runs on the bottom. Similarly, projections through the $y$-axis are shown in Figures \ref{fig:projFigs1} and \ref{fig:projFigs2}. It is clear that, regardless of the orientation of the magnetic fields, radiative cooling enhances the amount of dense gas in the cloud core. These dense cores are more stable against instabilities and survive for longer times than their counterparts without cooling. In fact, for the $M_{\rm hot} = 3.5$ case modeled here, the radiative clouds take almost twice as long as the non-cooling clouds to reach the point at which 50\% of the cloud mass is left. The specifics of the impact on mass loss are discussed in Section \ref{sec:evolution}.

Radiative cooling allows the clouds to compress into dense cloudlets and remain intact roughly twice as long as clouds without cooling. This is true in all cases; the hydrodynamic runs as well as the MHD runs with both magnetic field orientations. With magnetic fields impacting the clouds' ability to compress, it is clear that they will also have an impact on cooling efficiency. This connection has been discussed before \citep{Fragile2005} emphasizing that the transverse fields will enhance cloud compression and increase the cooling efficiency, while aligned fields will have the opposite effect by inhibiting compression. While we do see increased compression in the transverse field cases, this does not necessary translate to dense structures that survive over longer timescales than the other runs. A direct comparison to \citet{Fragile2005} is difficult to make as their simulations were in two dimensions and make use of a different cooling floor. However, our results are in qualitative agreement with their conclusions that radiative cooling extends cloud lifetimes while magnetic fields can either enhance or resist compression depending on the field orientation.

\subsection{Influence of Aligned Fields}

The disruption and morphology of the cloud differs significantly between runs H-rad-hr and A-rad-hr, which are shown in in the bottom left and center panels, respectively, of Figures \ref{fig:adiFigs1} and \ref{fig:adiFigs2}. The cloud in A-rad-hr (center) is compared to H-rad-hr (left) showing slices of the cloud density and $\beta$ at  2 $t_{\rm cc}$ (Figure \ref{fig:adiFigs1}) and 4 $t_{\rm cc}$ (Figure \ref{fig:adiFigs2}). The cloud within the magnetized wind is compressed at early times much like H-rad-hr, however the tail downwind of the cloud appears smoother in the MHD case. This is expected as strong magnetic fields aligned to the flow have been shown to inhibit the growth of Kelvin-Helmholtz instabilities \citep{Chandrabook, BandaBarragan2016}. In general, this creates a tail of cloud material flowing behind the cloud that is much less turbulent than the tail in H-rad-hr.

\begin{figure}
\includegraphics[angle=0,width=\columnwidth]{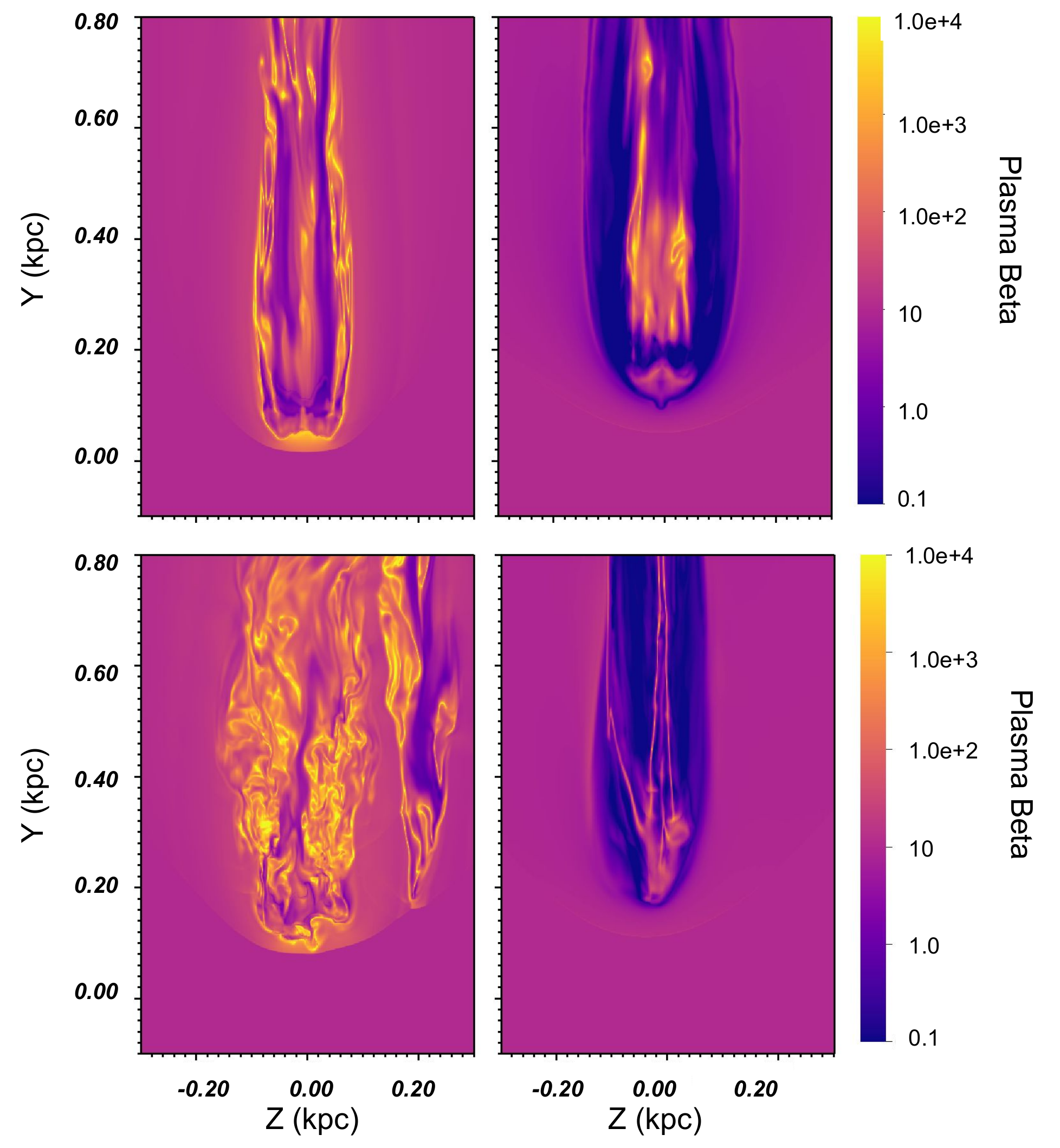}
\caption{Slices along the x-axis of plasma $\beta$ in A-rad-hr (left) and T-rad-hr (right) at 2 $t_{\rm cc}$ (top) and 4 $t_{\rm cc}$ (bottom). All lengths are given in kpc.
\label{fig:BetaFig}}
\end{figure}

These elongated tails in A-rad-hr are also regions where the magnetic pressure is comparable to the thermal pressure ($\beta \sim 1$) as shown in the left panels in Figure \ref{fig:BetaFig}. These tails are similar to the``flux rope" first described by \citet{MacLow1994}. In that study, field lines are pulled by the shock. As the wind passes the back of the cloud, surrounding gas fills in the space left by the higher velocity post-shock gas. This filling-in effect works to compress the field lines, resulting in an amplification of the magnetic field. Here we see the same amplification, with a similar structure to the ropes observed in \citet{Shin2008}.

These regions of amplified magnetic field also lead to notable differences in the evolution at later times. While H-rad-hr results in a few dense cloudlets that are slowly peeled away, the cloud in A-rad-hr is much more expanded and breaks up abruptly shortly after 5 $t_{\rm cc}$ with most of the cloud material evolving into lower-density wisps of gas. This is primarily driven by the magnetic pressure increasing faster than the thermal pressure with compression within the tail of the cloud. Thermal pressure is inversely proportional to volume, $P_{\rm th} \propto R^{-3},$
while magnetic pressure scales with radius as, $P_{\rm mag} \propto R^{-4}.$
This results in magnetic pressure in the MHD run opposing the compression to a greater extent than thermal pressure in the hydro case. This is most important for the material between the flux ropes. While the amplified ropes are created by the compression of converging flows, the intermediate material between these flux ropes is kept from condensing, resulting in more wispy fragments. These fragments and filaments are comparable to the structures seen in other studies \citep[e.g.][]{Fragile2005, Shin2008}. It is also worth noting the filament to the right, and the apparent asymmetry in the aligned field case at 4 t$_{\rm cc}$ in Figure 2 is likely caused by the amplification of tiny numerical differences, due to the growth of instabilities.

The inclusion of radiative cooling has the same effect on clouds embedded in aligned fields as it does in the hydrodynamic case. The cloud condenses into a dense core which then takes more time to be pulled apart by instabilities. The aligned fields may also aid in condensation as discussed in \citet{Gronke2018}. However, our domains do not extend far enough downwind to make a direct comparison. \citet{Gronke2018} find condensation at lengths $\approx 40 r_{\rm c}$ to $\approx 250 r_{\rm c}$ downwind of the cloud while our domain only extends to $\approx 13 r_{\rm c}$.

\subsection{Influence of Transverse Fields}\label{sec:trans}

We next consider the wind-cloud interaction in the case with transverse fields. Without magnetic fields, the reflected shock works to tear apart the cloud. However, in the case of a transverse field, this reflected shock is not created. Slices of the cloud density for A-rad-hr (bottom center) and T-rad-hr (bottom right) are shown in Figures \ref{fig:adiFigs1} and \ref{fig:adiFigs2}. The transverse fields produce a smooth, laminar flow with reduced effects of Kelvin-Helmhotlz (KH) instabilities due to the reorienting of the field lines as the wind pulls the lines to be more aligned with the flow. This is similar to what has been observed in previous studies \citep{Orlando2008, BandaBarragan2016, Gronnow2017, Gronnow2018}. In the right panels of Figure \ref{fig:BetaFig}, slices of $\beta$ are shown for T-rad-hr. Most of the cloud material is surrounded by an envelope of gas with $\beta \approx 1$. Here the thermal pressure and magnetic pressure are approximately equal, with the magnetic pressure providing resistance against the shock completely passing through the cloud.

This process of reorienting the fields lines is known as magnetic `draping', and it is an effective mechanism to shield dense gas from the erosive effects of dynamical instabilities
as shown in  \cite{Dursi2008, BandaBarragan2016}. These authors highlighted the potential of this effect to protect the cloud, increasing its stability and lifetime. While we also see evidence that magnetic draping  suppresses instabilities, we find that it does not ultimately increase the longevity of the cloud. In fact, the re-orientation of the field lines leads to another effect that causes cloud mass to be lost more quickly than the disruption from instabilities seen in the aligned and hydrodynamic cases.

\begin{figure}
\includegraphics[angle=0,width=\columnwidth]{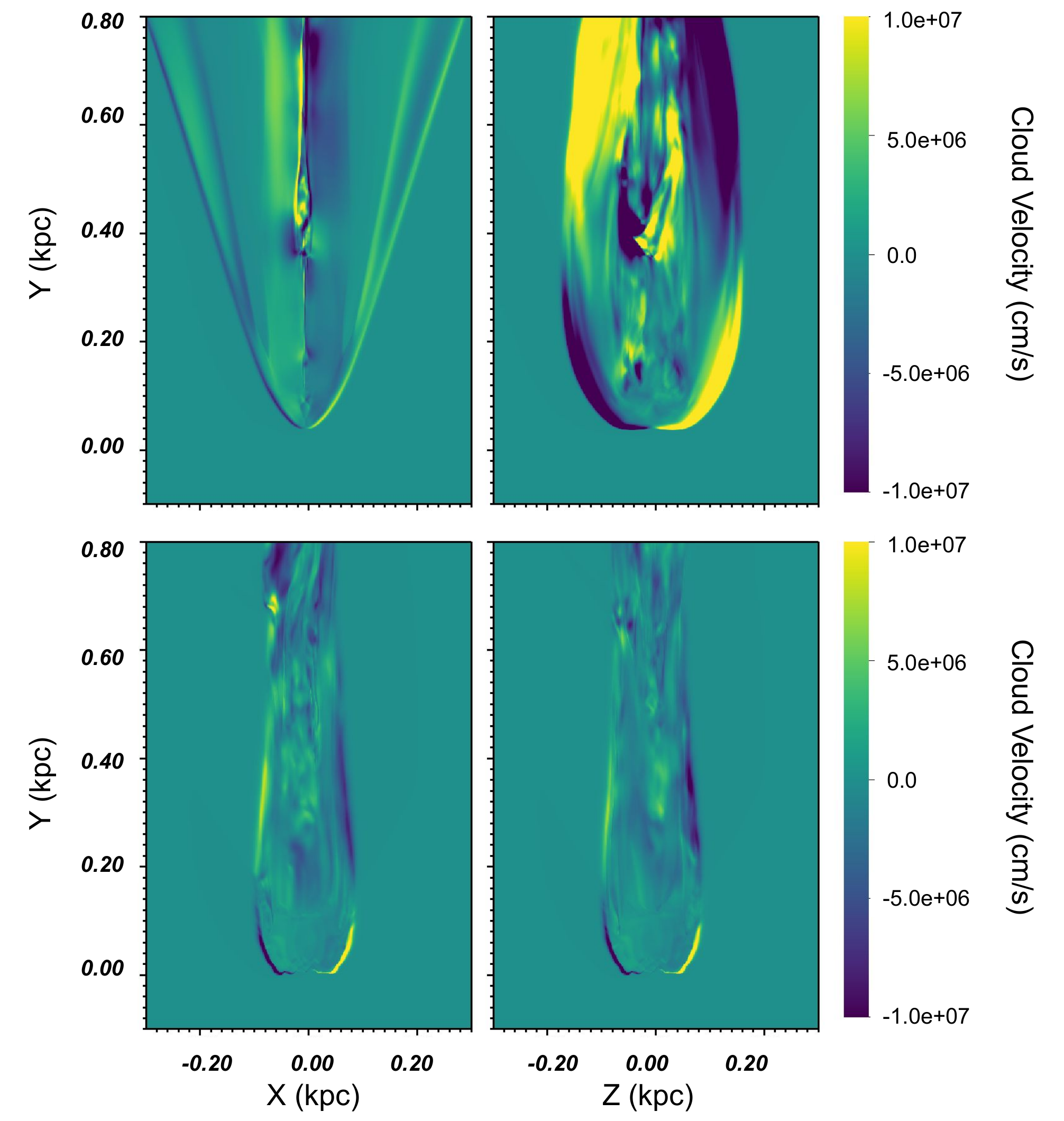}
\caption{Slices of the cloud velocity perpendicular to the wind [$x$ left; $z$ right] comparing T-rad-hr (top), and A-rad-hr (bottom) at 1.4 $t_{\rm cc}$. While the flow in A-rad-hr is symmetric, T-rad-hr preferentially flows in the $z$ direction around the core of the cloud.
\label{fig:VelCompare}}
\end{figure}

In the draping case, the magnetic field lines are pulled up sharply with the wind, causing an increase in magnetic pressure which pushes cloud material in the only free direction, the $z$-direction. In Figure \ref{fig:VelCompare}, the velocity of the cloud perpendicular to the wind is shown for T-rad-hr and A-rad-hr. In A-rad-hr, there is a symmetry between the $x$- and $z$-velocities with material primarily flowing only far enough to get around the leading edge of the cloud. For T-rad-hr, the cloud preferentially flows in the $z$-direction; much further than the original leading edge of the cloud, and at higher speeds than the A-hr material. This asymmetry is similar to that observed in previous studies \citep[i.e.][]{Gregori1999, McCourt2015, Gronnow2017}.

In particular, our results are in agreement with \citet{Gregori1999}, who shows the asymmetry produced by the expansion in the direction orthogonal to both the wind and field orientation in transverse field scenarios. \citet{Gregori1999, Gregori2000} also describe the role of Rayleigh-Taylor (RT) instabilities in forming a C-like structure at later times. The effects of these instabilities are amplified as the field lines become trapped and tangled at the front of the cloud \citep{BandaBarragan2016, Gronnow2017} . With trapped field lines the timescale of the growth of the RT instabilities is shortened, causing the front of the cloud in the simulations with transverse fields to be torn apart faster than in the aligned field or non-magnetized simulations. These amplified instabilities are responsible for the finger-like filaments seen in Figure \ref{fig:adiFigs2}.

\begin{figure*}
\includegraphics[angle=0, width=\textwidth]{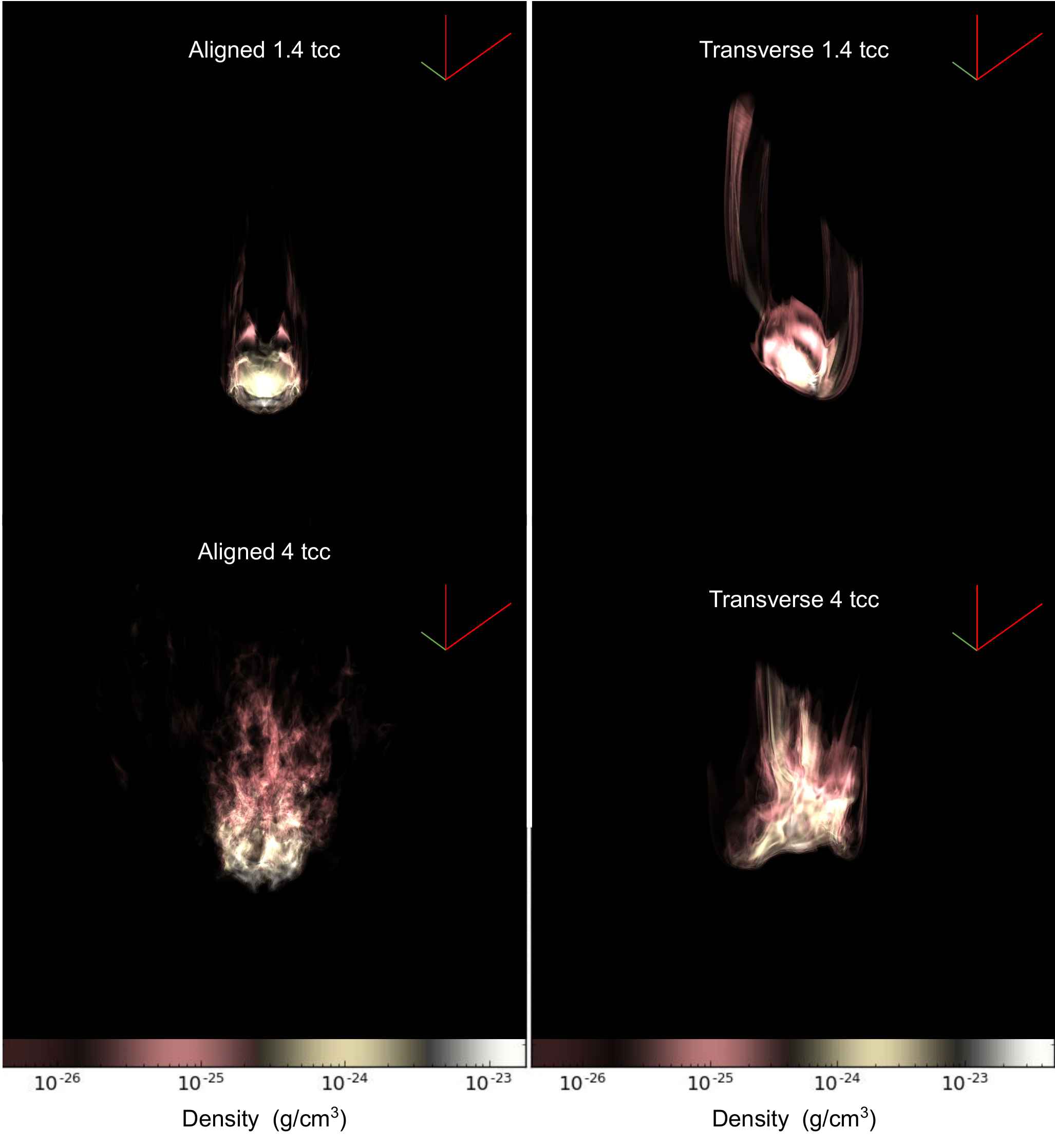}
\caption{Volume renderings of density for A-rad-hr and T-rad-hr at 1.4 $t_{\rm cc}$ and 4 $t_{\rm cc}$. There is clear asymmetry in T-rad-hr, which is flattened in the direction perpendicular to both the flow and magnetic fields.
\label{fig:volume}}
\end{figure*}

The squeezing of the cloud by the field lines produces a cloud with a more flattened appearance along the direction perpendicular to the wind and magnetic field lines (see the right-hand side panels of Figures \ref{fig:projFigs1} and \ref{fig:projFigs2}. This shape is also shown in the volume rendering in Figure \ref{fig:volume}. Compared to the cloud within the aligned field, the cloud within the transverse field maintains the smooth flow of mass down wind. At early times T-rad-hr is flattened and flowing around the core in the z direction while A-rad-hr appears symmetric with a more bullet-like shape. At later times T-rad-hr maintains this flattened shape as more material flows off of the core. A-rad-hr is more turbulent as material is being torn away by dampened, but present, hydrodynamic instabilities. This flattening of the cloud is in agreement with \citet{Shin2008} in which similar simulations produced sheet-like clouds parallel to the post-shock magnetic fields. Again, the apparent asymmetry at late times is likely caused by the amplification of tiny numerical differences (floating-point differences), due to the growth of linear instabilities.

\subsection{Strong Fields}\label{sec:strong}

\begin{figure}
\includegraphics[angle=0,width=\columnwidth]{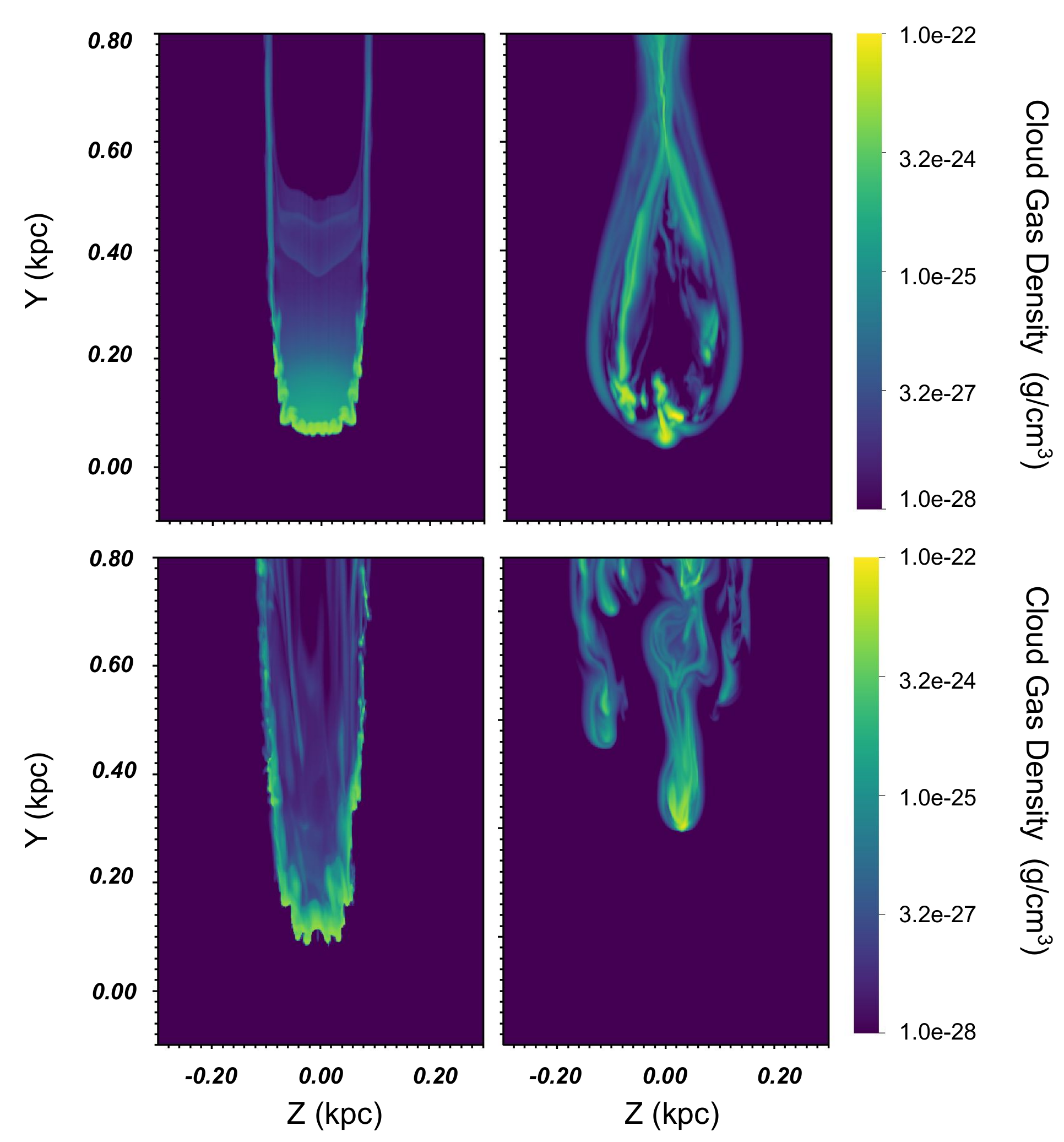}
\caption{Slices along the x-axis of cloud density comparing the strong field runs, A-B1-rad-hr (left) and T-B1-rad-hr (right) at 2 t$_{\rm cc}$ (top) and 4 t$_{\rm cc}$ (bottom). All densities are given in g/cm$^{-3}$ and all lengths are given in kpc. These are zoomed-in images of the more extended computational domains.
\label{fig:strongSlice}}
\end{figure}

\begin{figure}
\includegraphics[angle=0,width=\columnwidth]{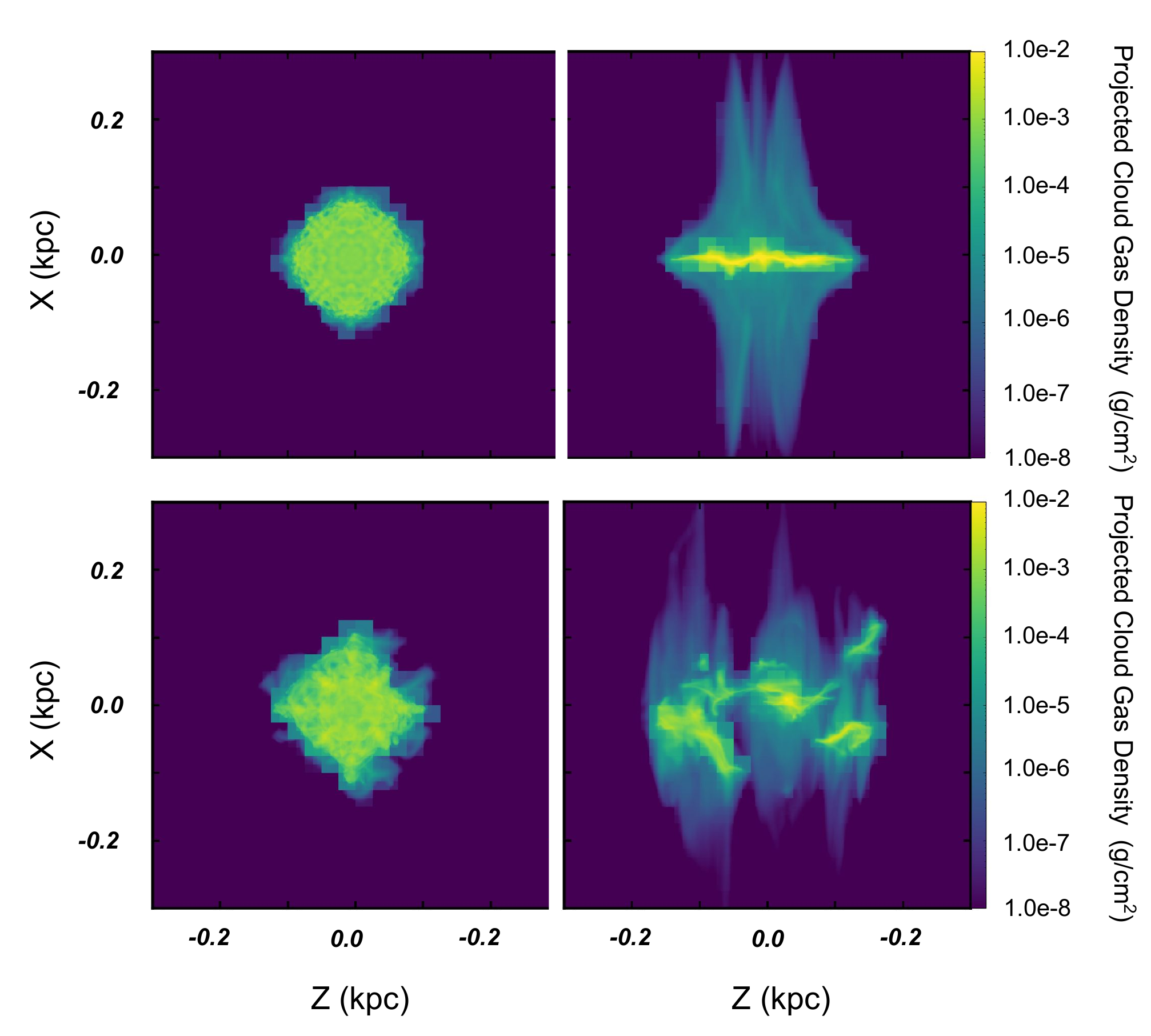}
\caption{Projections through the y-axis of cloud density comparing the strong field runs, A-B1-rad-hr (left) and T-B1-rad-hr (right) at 2 t$_{\rm cc}$ (top) and 4 t$_{\rm cc}$ (bottom). All column densities are given in g/cm$^{-2}$ and all lengths are given in kpc. These are zoomed-in images of the more extended computational domains. The low-resolution boundaries are due to the projection maintaining the resolution along the line of sight, which is dependent on the structure of the adaptive grid.
\label{fig:strongProj}}
\end{figure}

In addition to studying the effect of field orientation, we have investigated the effects of a stronger field. We consider each orientation, aligned and transverse, with an initial $\beta = 1$, this results in a field $\sim 3$ times stronger than the $\beta = 10$ cases. Slices and projections of the cloud density in these two strong field runs are shown in Figures \ref{fig:strongSlice} and \ref{fig:strongProj}.

In A-B1-rad-hr, the aligned strong fields lead to a significantly denser core than in A-rad-hr. However, as noted in previous studies \citep{Fragile2005}, the strong field suppresses low-temperature cooling, keeping the cloud from forming cloudlets. The main body of the cloud remains smooth as the KH instabilities are suppressed, leaving RT instabilities as the primarily cause of destruction. The small `flux ropes' formed in the tail of the cloud are no longer present, as the tail behind the cloud is made up of cold, low-density gas causing the entire tail of the cloud to have $\beta \sim 1$. At late times the cloud remains confined to a single core.

The squeezing effect seen in T-rad-hr is also apparent in T-B1-rad-hr. However, the destructive effects of the RT instabilities are further amplified with the stronger field causing the cloud to be torn apart from the front much faster. The cloudlets formed in this destruction phase are denser in T-B1-rad-hr than in T-rad-hr, but only up to an order or magnitude, consistent with results in \citet{Johansson2013}. As the main cloud is separated into smaller cloudlets, the rapid mass loss is exaggerated. Rather than stabilizing the cloud to allow a longer lifetime, the strong field results in destruction on time scales similar to the non-radiative clouds. 

\subsection{Evolution}\label{sec:evolution}

The morphology of these clouds has a significant impact on their overall evolution. In this section, we consider the evolution of the clouds in three global quantities; cloud mass loss, mixing fraction, and cloud velocity.

\begin{figure}
\includegraphics[angle=0,width=\columnwidth]{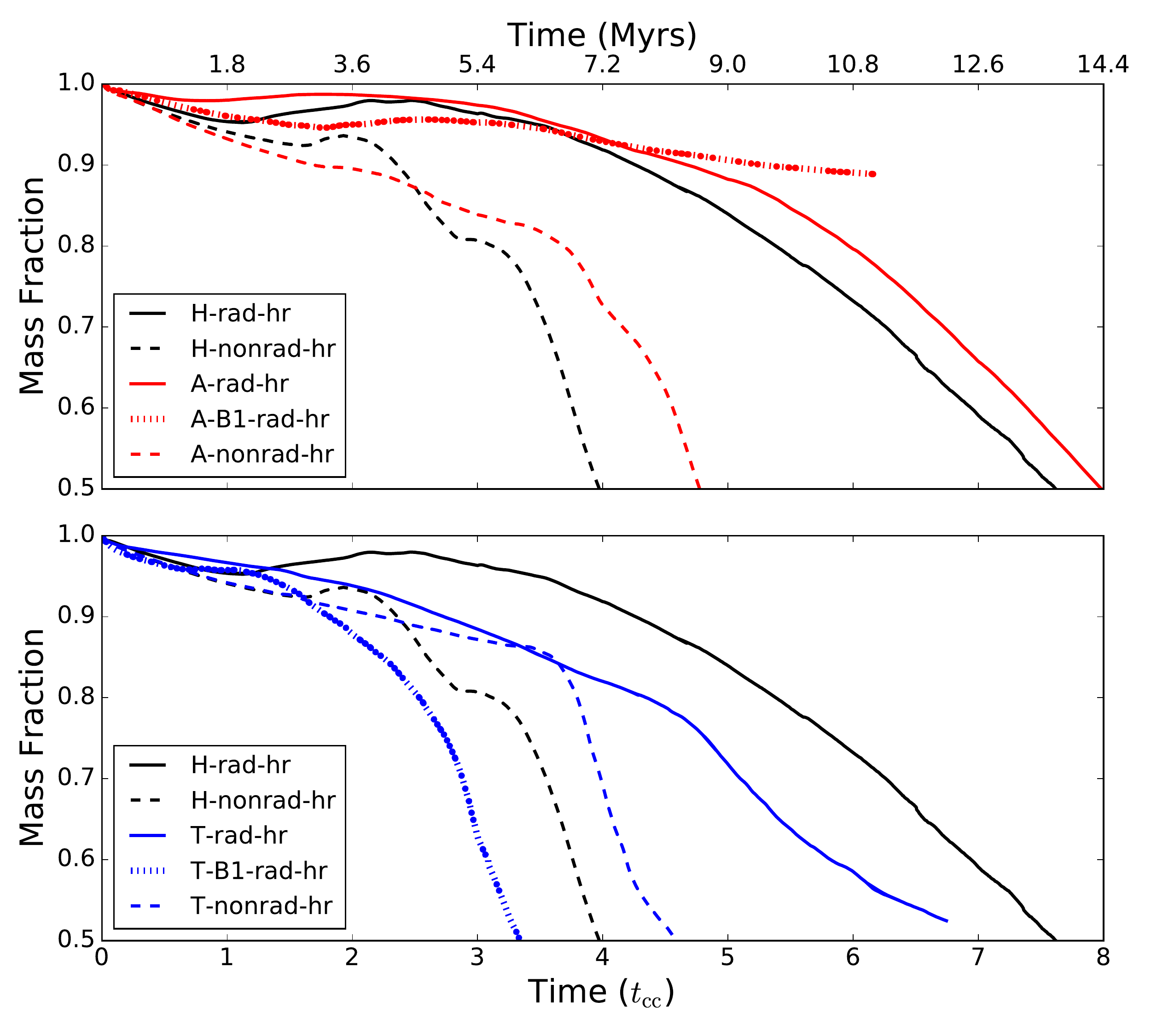}
\caption{Mass fraction of cloud material greater than $\rho_{\rm i, c}/3$ as a function of time in units of cloud crushing times of all high resolution runs. The hydro only simulations are shown in black; the cloud without cooling in a dashed line and the radiative cloud in a solid line. The aligned and transverse fields follow the same pattern in red and blue, respectively. Strong field runs are shown with dotted lines. 
\label{fig:massloss1}}
\end{figure}

In Figure \ref{fig:massloss1}, the fraction of cloud mass with density $ > \rho_{\rm c, i}/3$ is shown as a function of time for the hydrodynamic and MHD runs. Shown with the solid black line, the radiative hydrodynamic run follows the same mass loss rate as discussed in Paper I, with the fraction of remaining cloud mass staying above 90\% throughout the initial stages of the interaction before dropping as the cloud is destroyed by the wind at later stages.

Most notably, the transverse fields (solid blue line) do not appear to prolong the lifetime of the cloud. Rather than retaining a higher mass fraction for the majority of the simulation, T-rad-hr does the opposite. The mass fraction for the cloud with transverse fields decreases almost linearly for the first few cloud crushing times. While magnetic draping does somewhat protect the core of the cloud from shear instabilities, the bent field lines create inward magnetic forces that squeeze the cloud along the field direction and expand it in the perpendicular direction. This produces continuous mass loss as cloud material is carried with the wind. In the protective region that surrounds the cloud, the magnetic field has been amplified to 10 times the strength of the thermal pressure; 100 times greater than the initial magnetic pressure. At later times, the mass loss begins to increase as this region becomes thinner and the cloud material has been reduced to a long thin filament more vulnerable to instabilities. T-B1-rad-hr follows a similar evolution as the squeezing effect causes drastic mass loss at early times. However, the cloud T-B1-rad-hr is quickly torn apart by RT instabilities as the field is tangled in front of the cloud. This leads to the very steady mass loss past 2.5 t$_{\rm cc}$.

For the aligned fields in A-rad-hr (solid red line), the fields make little impact on the overall evolution, but they do lead to the abrupt break up of the cloud shortly after 5 t$_{\rm cc}$. Even though the KH short-wavelength instabilities are suppressed, the aligned fields only slightly improve the stability over H-rad-hr throughout the whole simulation. Since A-rad-hr does not form the same dense cloudlets as H-rad-hr, the extra mass comes from the `puffy' intermediate gas which breaks up from the main cloud to form filaments and wisps. This material, protected in high magnetic pressure bubbles, remains in the domain longer than material torn off the cloud in H-rad-hr. The mass loss for A-B1-rad-hr is similarly slow. The cloud remains in a single core as the suppressed instabilities are unable to pull it part and cause mass loss through ablation.

The evolution of the clouds without radiative cooling is distinctly different than those with radiative cooling. Curves for the mass loss for the runs without cooling are also shown in Figure \ref{fig:massloss1}. At early times, the mass loss for the magnetized runs is either on par with (transverse) or more significant (aligned) than the hydrodynamic run. From this perspective it may seem that fields do not increase cloud survival. However, at later times the ultimate destruction of the cloud occurs slightly sooner for H-nonrad-hr than either of the MHD runs without cooling. This indicates that while magnetic fields can impact cloud evolution in both the non-cooling and radiative cases, it is the combination of the fields and cooling that must be considered to predict the ultimate fate of the cloud.

As discussed in Section \ref{sec:trans}, the steep mass loss rate for the transverse field runs is due to mass being squeezed around the cloud as the magnetic pressure increases. The core of the cloud is embedded in a region of amplified fields with high magnetic pressure. In this region, there is little mixed material resulting in inefficient cooling rates. Due to this, the contribution to stability that radiative cooling provides in the other cases does not influence the primary mechanism for mass loss. It is not until the point that the cloud has become a filament ($\approx$ 4 $t_{\rm cc}$), with material breaking off through the draping layer does cooling begin to impact the mass loss. In T-nonrad-hr, the more exposed filament is unable to condense and begins to be torn apart causing a sharp decrease in mass. In T-rad-hr the filament is able to achieve a denser structure leading to the mass falling off at a slower rate, as seen in Figure \ref{fig:massloss1}.

Given that the A-rad-hr follows H-rad-hr much more closely than A-nonrad-hr, we conclude that the inclusion of cooling in the aligned MHD run has the same effect as it does in the hydrodynamic case.  In both cases, it aids in the compression of the cloud, leading to higher core densities and longer lifetimes as well as aiding the condensation of warm gas. The aligned magnetic fields provide a resistance to compression in the tail of the cloud which leads to expansion and break up, however they do not inhibit the effect of cooling to stabilize the cloud.

\begin{figure}
\includegraphics[angle=0,width=\columnwidth]{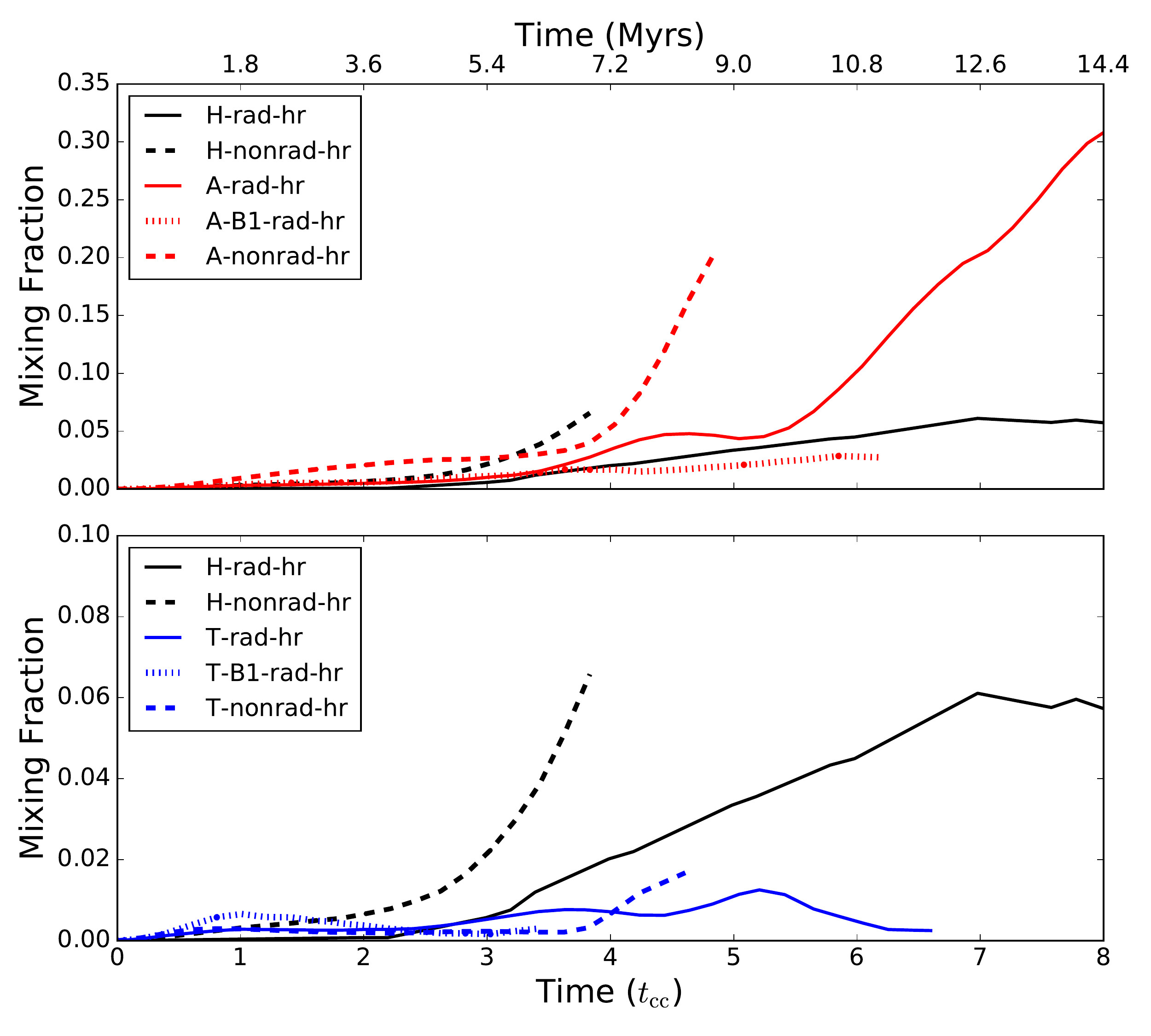}
\caption{Fraction of mixed material over time for both the no-cooling and radiative high resolution runs. The hydrodynamic simulations are shown in black; H-rad-hr as the solid line, H-nonrad-hr as the dashed line. The aligned and transverse fields follow the same pattern in red and blue, respectively. Strong field runs are shown with dotted lines. 
\label{fig:mixing_cool}}
\end{figure}

In Figure \ref{fig:mixing_cool} we show the mixing fraction as described in \citet{Xu1995} and \citet{Orlando2005},
\begin{equation}
f_{\rm mix} = \frac{1}{m_{\rm cloud, 0}} \int_{\rm (0.1 < C_{\rm cloud} < 0.9)} dV \rho C_{\rm cloud} ,
\end{equation}
where $m_{\rm cloud, 0}$ is the initial cloud mass and the integral is computed over the volume in which the tracer $C_{\rm cloud}$ is between 0.1 and 0.9. It is clear that the magnetic fields impact the mixing of material. As discussed in the previous section, the magnetic pressure in A-rad-hr keeps cloud material within the tail of the cloud, which allows the cloud material the opportunity to become mixed with the wind. This is reflected in very high mixing fractions as compared to the other two simulations, especially at later times where larger amounts of cloud material exist in a puffy intermediate phase after the clouds break up. However, A-B1-rad-hr does not have this same sharp increase in mixed material. As hydrodynamical instabilities are the mechanism for mixing, the stronger field case results in decreased mixing in relation to the decrease in disruption by the instabilities. This trend has been observed in previous studies \citep{Orlando2008}. In contrast, the transverse fields lead to very little mixing between the wind and cloud phases, resulting in mixing fractions even lower than H-rad-hr. The protection from the $\beta = 1$ envelope in both T-rad-hr and T-B1-rad-hr effectively confines the cloud material, restricting the possibility of mixing.

For the cases without cooling, the clouds are not able to condense and they become well mixed with the wind in H-nonrad-hr and A-nonrad-hr. The mixing in these clouds increases at later times as the cloud begins to be torn apart by instabilities. Were the simulation to continue well past the point where 50\% of the cloud mass was left, the fraction of mixed material would continue to increase as the cloud ablates and drifts downwind. In contrast, for the transverse run without cooling, the $\beta = 1$ envelope is still an effective form of protection, keeping  the mixing fraction well below 0.01.

\begin{figure}
\includegraphics[angle=0,width=\columnwidth]{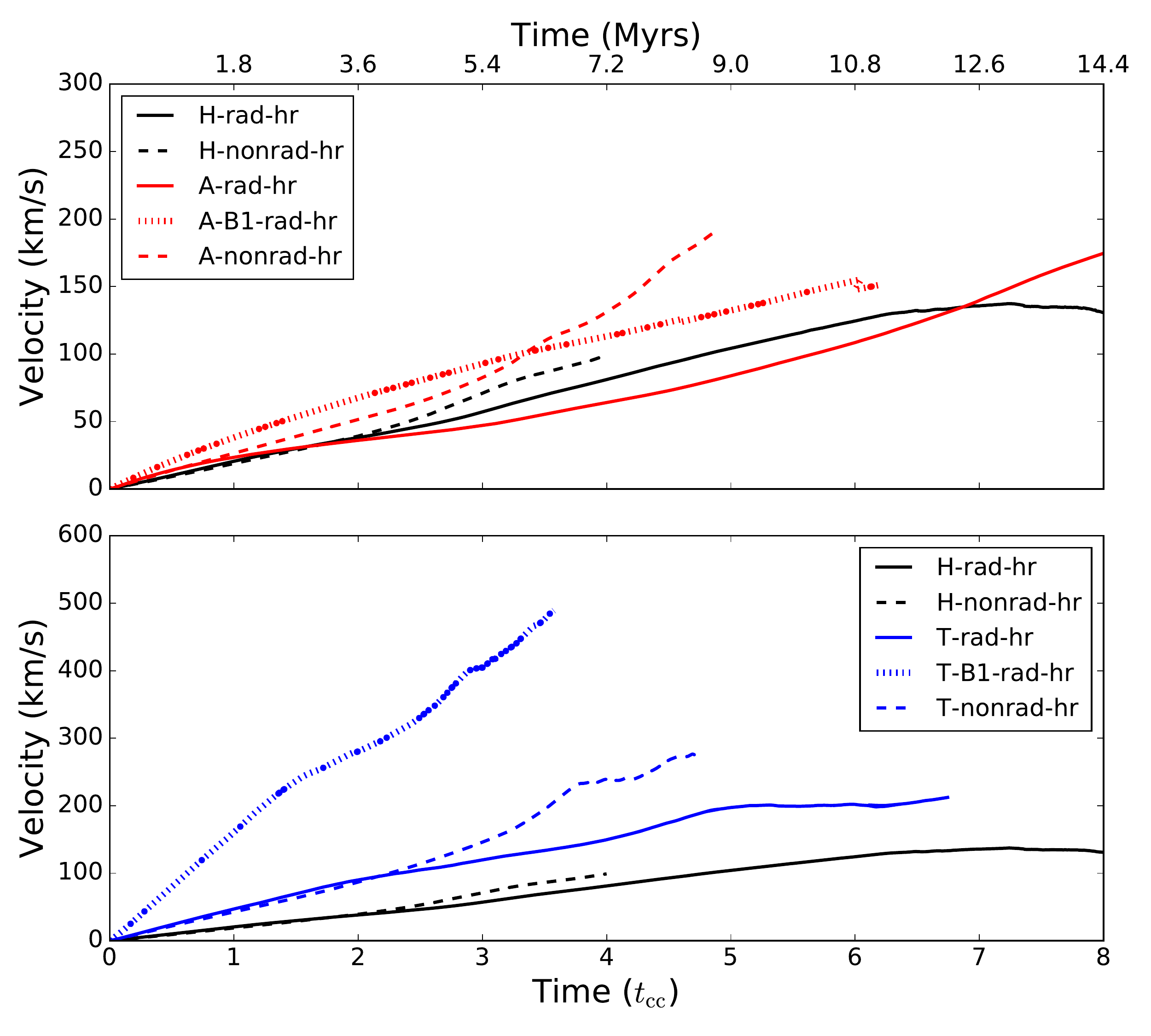}
\caption{Cloud velocity along the flow of the wind as a function of time in units of cloud crushing times of all high resolution runs. The hydrodynamic simulations are shown in black; H-rad-hr as the solid line, H-nonrad-hr as the dashed line. The aligned and transverse fields shown with the same pattern in red and blue respectively. Strong field runs are shown with dotted lines. 
\label{fig:CompareVel_Cool}}
\end{figure}

Finally, we consider the impact of magnetic fields on the acceleration of the cloud. In Figure \ref{fig:CompareVel_Cool}, the average down-wind velocity of the cloud is shown with time. In the absence of magnetic fields, the cloud is accelerated consistently. For the aligned fields, due to the fact that the magnetic pressure is not in the direction of the acceleration, there is little difference between A-rad-hr and H-rad-hr \citep[in agreement with ][]{Jones1996, MacLow1994}. The acceleration of the cloud within the transverse fields is much higher than either of the other two runs. From the same magnetic pressure argument, as the transverse field lines are compacted by the flow at the front of the cloud, this leads to an amplification in the magnetic field corresponding to an increase in magnetic pressure. At the leading edge of the cloud, just inside the $\beta = 1$ envelope, the magnetic pressure has been amplified to 100 times the initial pressure. This pressure is at the leading edge of the cloud, pushing in the same direction as the ram pressure acceleration. Due to this additional pressure, the cloud has a larger acceleration in the T-rad-hr run than in the other two runs. This effect is even more apparent in T-B1-rad-hr, where the stronger field leads to an even higher magnetic pressure at the front of the cloud which accelerates the cloud three times faster than in T-rad-hr, in agreement with the increase in initial magnetic field strength from an initial $\beta = 10$ (1.86 $\mu$G) to an initial $\beta = 1$ (5.88 $\mu$G). In the runs without cooling, the velocity of the cloud increases similarly to the radiative cases at early times. At later times, without the ability to cool and condense, these clouds are torn apart and accelerated to higher velocities.

The evolutionary trends for these runs can easily be summarized as follows. As most destruction is through forces perpendicular to the wind flow, aligned fields have little to no impact on the mass loss and cloud velocity. However, the additional pressure the magnetic fields provide leads to expansion, break up and higher amounts of cloud mass intermixed with wind material. Conversely, transverse fields lead to increased acceleration and larger amounts of poorly-mixed cloud material being lost from the domain. These effects are both due to the amplification and draping of field lines as they are dragged with the flow of the wind, which also leads to reduced mixing between the cloud and wind materials.

\subsection{Resolution Effects and Limitations}

In Paper I, we discussed the resolution effects on these hydrodynamic simulations with radiative cooling. Considering the same low and fiducial resolutions as in the current paper, $\Delta x = R_{\rm cloud}/32$ and $\Delta x = R_{\rm cloud}/64$ , as well as a high resolution run with $\Delta x = R_{\rm cloud}/128$, we highlighted that the under-resolved instabilities in the  $\Delta x = R_{\rm cloud}/32$ significantly impact the resulting mass loss estimates. The high-resolution, $\Delta x = R_{\rm cloud}/128$ run, on the other hand, converged to the same solution for mass loss as the fiducial run, but it also captured more diffuse material, leading to higher mixing fractions.

\begin{figure}
\includegraphics[angle=0,width=\columnwidth]{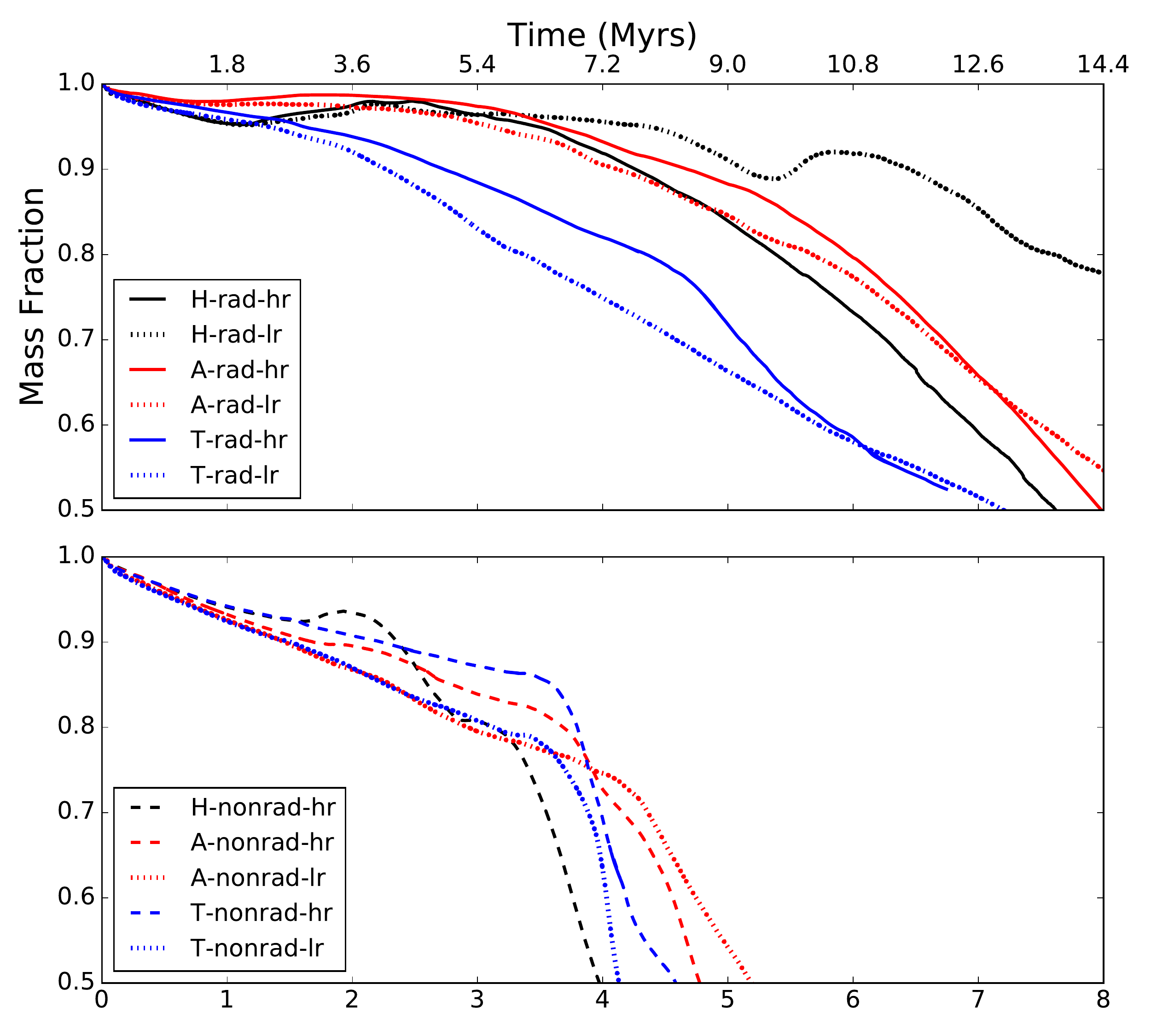}
\caption{Top Panel: Mass fraction of cloud material greater than $\rho_{\rm i, c}/3$ as a function of time in units of cloud crushing times of all runs. The hydrodynamic simulations are shown in black; H-rad-hr as the solid line, H-rad-lr as the dotted line. The aligned and transverse fields follow the same pattern in red and blue, respectively. Bottom Panel: Same as the top panel comparing the high and low resolution runs without radiative cooling, dashed lines are high resolution while dotted lines are low resolution.
\label{fig:massloss2}}
\end{figure}

Taking the same approach, we compare the mass loss, mixing fractions and cloud velocity between the high-resolution, $\Delta x = R_{\rm cloud}/64$ and low-resolution $\Delta x = R_{\rm cloud}/32$ runs with and without magnetic fields. In the top panel of Figure \ref{fig:massloss2} the mass loss of H-rad-lr is much lower than all other runs, deviating significantly by 4 $t_{\rm cc}$, the same point at which A-rad-hr and A-rad-lr begin to depart from H-rad-hr. This time corresponds to the transition between a single shocked cloud core to several smaller cloud cores as the cloud begins to break up. Unlike the hydro cases, runs A-rad-hr and A-rad-lr are similar to each other, and lie between the two estimates from the hydro simulations. Thus it is clear that convergence properties of the aligned cases are better than the hydro cases, and that the $\Delta x = R_{\rm cloud}/64$ resolution of H-rad-hr is sufficient to conclude that aligned magnetic fields slightly decrease the cloud mass loss rate. Finally, in the transverse field runs, the coarse resolution in T-rad-lr leads to more mass loss than T-rad-hr, but the resolution effects are again smaller than in the hydro runs.

\begin{figure}
\includegraphics[angle=0,width=\columnwidth]{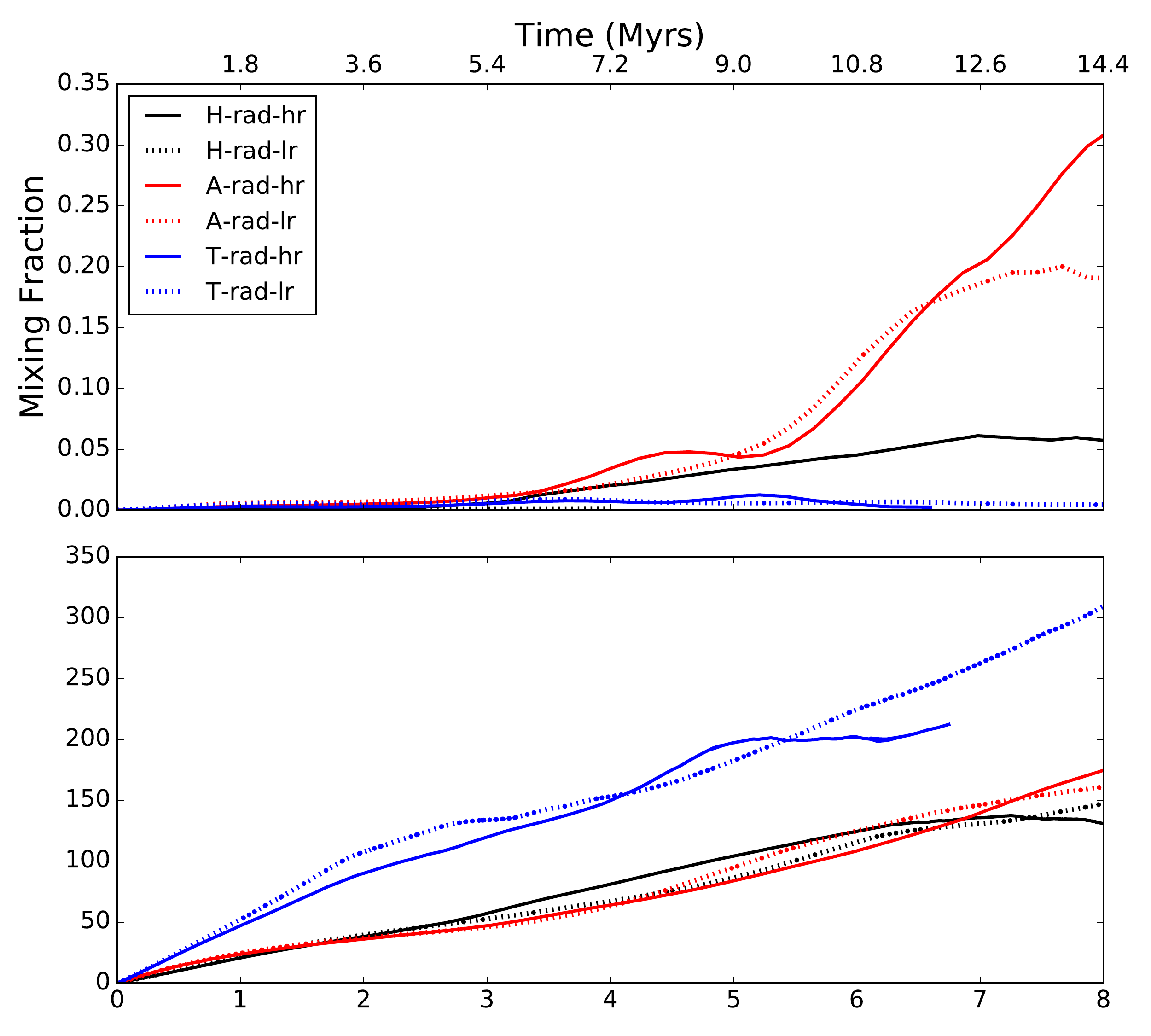}
\caption{Top panel: Fraction of mixed material over time comparing resolution. Bottom panel: Cloud velocity along the flow of the wind as a function of time in units of cloud crushing times of all runs. The hydrodynamic simulations are shown in black; H-rad-hr as the solid line, H-rad-lr as the dotted line. The aligned and transverse fields follow the same pattern in red and blue, respectively.
\label{fig:mix_vel}}
\end{figure}

In the top panel of Figure \ref{fig:mix_vel} we show the effect on resolution of the mixing fraction.
As discussed in the previous section, the magnetic pressure in A-rad-hr keeps cloud material within the tail of the cloud. While A-rad-hr follows H-rad-hr at early times, past 5 $t_{\rm cc}$ the mixing fraction for the magnetized case increases to over three times that of the hydro case for both resolutions. The cloud material is kept within dense cloudlets in H-rad-hr and H-rad-lr while in A-rad-hr and A-rad-lr it becomes well mixed with the wind material. While it is clear that an increase in resolution leads to an increase in mixing fraction in both the hydrodynamic and aligned MHD cases, we can qualitatively conclude that the presence of aligned magnetic fields leads to more mixing overall. In contrast, the difference in the mixing fraction between the two transverse field runs is small. This confirms that there is limited mixing in these runs and higher levels of refinement do not reveal more intermediate material.

Finally, the velocity evolution of the radiative clouds across resolutions is shown in the lower panel of  Figure \ref{fig:mix_vel}. The two resolutions are consistent with each other over the duration of the simulations for both the hydrodynamic and MHD runs. This further enforces the argument that our resolution of $\Delta x = R_{\rm cloud}/64$ is sufficient.

In Paper I, we were able to show explicitly that going to a resolution of $\Delta x = R_{\rm cloud}/128$ gives similar results as the $\Delta x = R_{\rm cloud}/64$ simulations for the hydro-only cases, but it is computationally prohibitive to conduct similar simulations for the MHD case, as a single $R_{\rm c}/128$ MHD simulation would require over 100k node-hours on Stampede2 with 68 cores per node. With only two resolutions, it is not possible to conclude that these values converge monotonically. However, the evolution of mass, mixing fraction, and velocity, are much more consistent with each other between $R_{\rm c}/32$ and $R_{\rm c}/64$ than the hydrodynamic runs. This is true for both the non-cooling and radiative simulations, giving us confidence that our results have captured the overall evolution of radiative, magnetized clouds.

On the other hand, the choices for the magnetic field orientations in this work are idealized and do not fully reflect the more complex topologies of astrophysical fields. In reality, magnetic fields in the IGM are random and tangled. These components would likely create an additional stabilizing pressure (see citet{BandaBarragan2018}) which may ultimately affect the cooling efficiency of the clouds. However, the two choices for field orientations here capture the general cases that will influence realistic configurations.\citet{Gronnow2018,Gronke2018,Gronke2019} have shown that condensation can impact the cold gas within the interaction by creating more of the dense gas downwind. With a domain large enough to capture this condensed gas, we may find that the mass flowing around and behind the cloud in the transverse cases is not completely lost sustaining the colder cloud phase in the interaction for longer times.

In addition to the limitations imposed by domain size, our results are subject to numerical effects. While we have chosen our orientations to migate the effects, numerical resistivity can result in unphysical magnetic reconnection, particularly where field lines have been bent around the cloud by the wind. As this is a resolution-dependent effect, the choice of AMR refinement criteria can impact the location and scale of these effects.

Finally, we note that our results are also dependent on the choice of cooling regimes and cooling floor which limit the extent to which gas can cool and condense down wind. Our results are also limited by the exclusion of heating from UV radiation and cosmic rays. These factors may reduce the cloud's ability to cool and form dense cloudlets. Self-contained and turbulent fields as well as a smooth cloud density profile may also lead to different quantitative results. \citet{BandaBarragan2018} have started to explore the effects of turbulence in wind-cloud problems, but without radiative cooling. Thus, combining cooling and turbulence should be subject to a follow-up study.

\section{Summary and Conclusions} \label{sec:summary}

We have presented a suite of three-dimensional AMR MHD wind-cloud simulations including radiative cooling, and investigated the effect of magnetic fields in two orientations on the disruption and evolution of the wind-cloud interaction. Our conclusions can be summarized as follows:

\begin{enumerate}
\item Radiative cooling extends the lifetime of all clouds, regardless of whether or not magnetic fields are present and regardless of their particular orientation.
\item Magnetic fields aligned with the wind protect the cloud from hydrodynamic instabilities, creating a smoother cloud morphology, but they do not provide a substantial increase to the cloud's lifetime or stability over the non-magnetized case. The magnetic pressure resists compression in the tail of the cloud resulting in slightly more diffuse structures with higher mixing fractions after the clouds break up.
\item Clouds embedded in magnetic fields transverse to the wind experience a draping effect, which does not aid cloud survival if the flow is radiative and can cool. Instead, the amplified and re-oriented magnetic field in the wind pushes the cloud material in the direction perpendicular to the field, leading to higher rates of mass loss.
\item The magnetic draping that occurs with transverse magnetic fields allows magnetic and thermal pressures to reach equipartition. Thus, magnetic draping is an effective acceleration mechanism, as its effect becomes more significant in models with stronger transverse fields.
\item The protection of a $\beta = 1$ envelope prevents the cloud material draped by transverse fields from mixing with the wind, as it is contained in a region of high magnetic pressure that opposes ram pressure. Cooling is ineffective in this envelope and condensation is reduced.
\item An increase in field strength amplifies the effects of transverse fields, pulling the cloud apart at a faster rate. For algined fields, a stronger field strength results in an increase in cloud lifetime.

\item Magnetic fields inhibit small-scale hydrodynamic instabilities, so the two resolutions of the radiative MHD runs are in better agreement with each other than their hydrodynamic counterparts.
\end{enumerate}

Together these results demonstrate that the influence of magnetic fields has a significant impact on the evolution in wind-cloud interactions. These conclusions are applicable to the hot phase of galacitc winds and the general study of the interaction of magnetized clouds and hot winds. It is clear that radiative cooling always aids to extend cloud lifetime, however the combined effects of cooling and magnetic fields do not compound to produce more stable clouds. Instead, magnetic fields can be prohibitive to the stabilizing effects of radiative cooling. The distinction between these two effects is highly dependent on the orientation of the field with respect to the wind. Our results emphasize the need for studies to account for multiple physical effects simultaneously.  Investigating the role of magnetic fields in combination with effects such as turbulence, self-gravity and anisotropic conduction will improve our understanding of the multiphase nature of outflowing winds.

\acknowledgments
This work was supported by the National Science Foundation under grants OISE-1458445 and AST-1715876. MB is supported by a grant by the Deutsche Forschungsgemeinschaft under BR2026-12. WBB is supported by the Deutsche Forschungsgemeinschaft (DFG) via grant BR2026215. 

We would like to thank the Texas Advanced Computing Center (TACC) at The University of Texas at Austin, and the Extreme Science and Engineering Discovery Environment (XSEDE) for providing HPC resources via grants TG-AST130021 and TG-AST160063 that have contributed to the results reported within this paper. The software used in this work was in part developed by the DOE-supported Flash Center for Computational Science at the University of Chicago. C.~F.~acknowledges funding provided by the Australian Research Council (Discovery Project DP170100603 and Future Fellowship FT180100495), and the Australia-Germany Joint Research Cooperation Scheme (UA-DAAD).

\bibliography{draft_1_arxiv}

\begin{thebibliography}{64}
\expandafter\ifx\csname natexlab\endcsname\relax\def\natexlab#1{#1}\fi

\bibitem[{{Adebahr} {et~al.}(2017){Adebahr}, {Krause}, {Klein}, {Heald}, \&
  {Dettmar}}]{Adebahr2018}
{Adebahr}, B., {Krause}, M., {Klein}, U., {Heald}, G., \& {Dettmar}, R.-J.
  2017, \aap, 608, A29

\bibitem[{{Agertz} \& {Kravtsov}(2015)}]{Agertz2015}
{Agertz}, O. \& {Kravtsov}, A.~V. 2015, \apj, 804, 18

\bibitem[{{Banda-Barrag{\'a}n} {et~al.}(2018){Banda-Barrag{\'a}n}, {Federrath},
  {Crocker}, \& {Bicknell}}]{BandaBarragan2018}
{Banda-Barrag{\'a}n}, W.~E., {Federrath}, C., {Crocker}, R.~M., \& {Bicknell},
  G.~V. 2018, \mnras, 473, 3454

\bibitem[{{Banda-Barrag{\'a}n} {et~al.}(2016){Banda-Barrag{\'a}n}, {Parkin},
  {Federrath}, {Crocker}, \& {Bicknell}}]{BandaBarragan2016}
{Banda-Barrag{\'a}n}, W.~E., {Parkin}, E.~R., {Federrath}, C., {Crocker},
  R.~M., \& {Bicknell}, G.~V. 2016, \mnras, 455, 1309

\bibitem[{{Banda-Barrag{\'a}n} {et~al.}(2019){Banda-Barrag{\'a}n}, {Zertuche},
  {Federrath}, {Garc{\'\i}a Del Valle}, {Br{\"u}ggen}, \&
  {Wagner}}]{BandaBarragan2019}
{Banda-Barrag{\'a}n}, W.~E., {Zertuche}, F.~J., {Federrath}, C., {Garc{\'\i}a
  Del Valle}, J., {Br{\"u}ggen}, M., \& {Wagner}, A.~Y. 2019, \mnras, 486, 4526

\bibitem[{Berger \& Colella(1989)}]{Berger1989}
Berger, M. \& Colella, P. 1989, Journal of Computational Physics, 82, 64

\bibitem[{{Bolatto} {et~al.}(2013){Bolatto}, {Warren}, {Leroy}, {Walter},
  {Veilleux}, {Ostriker}, {Ott}, {Zwaan}, {Fisher}, {Weiss}, {Rosolowsky}, \&
  {Hodge}}]{Bolatto2013}
{Bolatto}, A.~D., {Warren}, S.~R., {Leroy}, A.~K., {Walter}, F., {Veilleux},
  S., {Ostriker}, E.~C., {Ott}, J., {Zwaan}, M., {Fisher}, D.~B., {Weiss}, A.,
  {Rosolowsky}, E., \& {Hodge}, J. 2013, \nat, 499, 450

\bibitem[{{Br{\"u}ggen} \& {Scannapieco}(2016)}]{Bruggen2016}
{Br{\"u}ggen}, M. \& {Scannapieco}, E. 2016, \apj, 822, 31

\bibitem[{{Carretti} {et~al.}(2013){Carretti}, {Crocker}, {Staveley-Smith},
  {Haverkorn}, {Purcell}, {Gaensler}, {Bernardi}, {Kesteven}, \&
  {Poppi}}]{Carretti2013}
{Carretti}, E., {Crocker}, R.~M., {Staveley-Smith}, L., {Haverkorn}, M.,
  {Purcell}, C., {Gaensler}, B.~M., {Bernardi}, G., {Kesteven}, M.~J., \&
  {Poppi}, S. 2013, \nat, 493, 66

\bibitem[{Chandrasekhar(1981)}]{Chandrabook}
Chandrasekhar, S. 1981, Hydrodynamic and Hydromagnetic Stability, Dover Books
  on Physics Series (Dover Publications)

\bibitem[{{Chevalier} \& {Clegg}(1985)}]{Chevalier1985}
{Chevalier}, R.~A. \& {Clegg}, A.~W. 1985, \nat, 317, 44

\bibitem[{{Colella} \& {Woodward}(1984)}]{Colella1984}
{Colella}, P. \& {Woodward}, P.~R. 1984, Journal of Computational Physics, 54,
  174

\bibitem[{{Cooper} {et~al.}(2009){Cooper}, {Bicknell}, {Sutherland}, \&
  {Bland-Hawthorn}}]{Cooper2009}
{Cooper}, J.~L., {Bicknell}, G.~V., {Sutherland}, R.~S., \& {Bland-Hawthorn},
  J. 2009, \apj, 703, 330

\bibitem[{{Dav{\'e}} {et~al.}(2011){Dav{\'e}}, {Finlator}, \&
  {Oppenheimer}}]{Dave2011}
{Dav{\'e}}, R., {Finlator}, K., \& {Oppenheimer}, B.~D. 2011, \mnras, 416, 1354

\bibitem[{{Dubey} {et~al.}(2008){Dubey}, {Reid}, \& {Fisher}}]{Dubey2008}
{Dubey}, A., {Reid}, L.~B., \& {Fisher}, R. 2008, Physica Scripta Volume T,
  132, 014046

\bibitem[{{Dursi} \& {Pfrommer}(2008)}]{Dursi2008}
{Dursi}, L.~J. \& {Pfrommer}, C. 2008, \apj, 677, 993

\bibitem[{{Fragile} {et~al.}(2005){Fragile}, {Anninos}, {Gustafson}, \&
  {Murray}}]{Fragile2005}
{Fragile}, P.~C., {Anninos}, P., {Gustafson}, K., \& {Murray}, S.~D. 2005,
  \apj, 619, 327

\bibitem[{{Fryxell} {et~al.}(2000){Fryxell}, {Olson}, {Ricker}, {Timmes},
  {Zingale}, {Lamb}, {MacNeice}, {Rosner}, {Truran}, \& {Tufo}}]{Fryxell2000}
{Fryxell}, B., {Olson}, K., {Ricker}, P., {Timmes}, F.~X., {Zingale}, M.,
  {Lamb}, D.~Q., {MacNeice}, P., {Rosner}, R., {Truran}, J.~W., \& {Tufo}, H.
  2000, \apjs, 131, 273

\bibitem[{{Gray} \& {Scannapieco}(2010)}]{Gray2010}
{Gray}, W.~J. \& {Scannapieco}, E. 2010, \apj, 718, 417

\bibitem[{{Gregori} {et~al.}(1999){Gregori}, {Miniati}, {Ryu}, \&
  {Jones}}]{Gregori1999}
{Gregori}, G., {Miniati}, F., {Ryu}, D., \& {Jones}, T.~W. 1999, \apjl, 527,
  L113

\bibitem[{{Gregori} {et~al.}(2000){Gregori}, {Miniati}, {Ryu}, \&
  {Jones}}]{Gregori2000}
---. 2000, \apj, 543, 775

\bibitem[{{Gronke} \& {Oh}(2018)}]{Gronke2018}
{Gronke}, M. \& {Oh}, S.~P. 2018, \mnras, 480, L111

\bibitem[{{Gronke} \& {Oh}(2019)}]{Gronke2019}
---. 2019, arXiv e-prints, arXiv:1907.04771

\bibitem[{{Gr{\o}nnow} {et~al.}(2018){Gr{\o}nnow}, {Tepper-Garc{\'{\i}}a}, \&
  {Bland-Hawthorn}}]{Gronnow2018}
{Gr{\o}nnow}, A., {Tepper-Garc{\'{\i}}a}, T., \& {Bland-Hawthorn}, J. 2018,
  \apj, 865, 64

\bibitem[{{Gr{\o}nnow} {et~al.}(2017){Gr{\o}nnow}, {Tepper-Garc{\'{\i}}a},
  {Bland-Hawthorn}, \& {McClure-Griffiths}}]{Gronnow2017}
{Gr{\o}nnow}, A., {Tepper-Garc{\'{\i}}a}, T., {Bland-Hawthorn}, J., \&
  {McClure-Griffiths}, N.~M. 2017, \apj, 845, 69

\bibitem[{{Johansson} \& {Ziegler}(2013)}]{Johansson2013}
{Johansson}, E. P.~G. \& {Ziegler}, U. 2013, \apj, 766, 45

\bibitem[{{Jones} {et~al.}(1996){Jones}, {Ryu}, \& {Tregillis}}]{Jones1996}
{Jones}, T.~W., {Ryu}, D., \& {Tregillis}, I.~L. 1996, \apj, 473, 365

\bibitem[{{Kacprzak} {et~al.}(2014){Kacprzak}, {Martin}, {Bouch{\'e}},
  {Churchill}, {Cooke}, {LeReun}, {Schroetter}, {Ho}, \&
  {Klimek}}]{Kacprzak2014}
{Kacprzak}, G.~G., {Martin}, C.~L., {Bouch{\'e}}, N., {Churchill}, C.~W.,
  {Cooke}, J., {LeReun}, A., {Schroetter}, I., {Ho}, S.~H., \& {Klimek}, E.
  2014, \apjl, 792, L12

\bibitem[{{Klein} {et~al.}(1994){Klein}, {McKee}, \& {Colella}}]{Klein1994}
{Klein}, R.~I., {McKee}, C.~F., \& {Colella}, P. 1994, \apj, 420, 213

\bibitem[{{Li} {et~al.}(2013){Li}, {Frank}, \& {Blackman}}]{Li2013}
{Li}, S., {Frank}, A., \& {Blackman}, E.~G. 2013, \apj, 774, 133

\bibitem[{{Li} {et~al.}(2020){Li}, {Hopkins}, {Squire}, \& {Hummels}}]{Li2020}
{Li}, Z., {Hopkins}, P.~F., {Squire}, J., \& {Hummels}, C. 2020, \mnras, 492,
  1841

\bibitem[{{Lu} {et~al.}(2015){Lu}, {Blanc}, \& {Benson}}]{Lu2015}
{Lu}, Y., {Blanc}, G.~A., \& {Benson}, A. 2015, \apj, 808, 129

\bibitem[{{Mac Low} \& {Ferrara}(1999)}]{MacLow1999}
{Mac Low}, M.-M. \& {Ferrara}, A. 1999, \apj, 513, 142

\bibitem[{{Mac Low} {et~al.}(1994){Mac Low}, {McKee}, {Klein}, {Stone}, \&
  {Norman}}]{MacLow1994}
{Mac Low}, M.-M., {McKee}, C.~F., {Klein}, R.~I., {Stone}, J.~M., \& {Norman},
  M.~L. 1994, \apj, 433, 757

\bibitem[{Marder(1987)}]{Marder1987}
Marder, B. 1987, Journal of Computational Physics, 68, 48

\bibitem[{{McClure-Griffiths} {et~al.}(2013){McClure-Griffiths}, {Green},
  {Hill}, {Lockman}, {Dickey}, {Gaensler}, \& {Green}}]{McClure2013}
{McClure-Griffiths}, N.~M., {Green}, J.~A., {Hill}, A.~S., {Lockman}, F.~J.,
  {Dickey}, J.~M., {Gaensler}, B.~M., \& {Green}, A.~J. 2013, \apjl, 770, L4

\bibitem[{{McCourt} {et~al.}(2018){McCourt}, {Oh}, {O'Leary}, \&
  {Madigan}}]{McCourt2018}
{McCourt}, M., {Oh}, S.~P., {O'Leary}, R., \& {Madigan}, A.-M. 2018, \mnras,
  473, 5407

\bibitem[{{McCourt} {et~al.}(2015){McCourt}, {O'Leary}, {Madigan}, \&
  {Quataert}}]{McCourt2015}
{McCourt}, M., {O'Leary}, R.~M., {Madigan}, A.-M., \& {Quataert}, E. 2015,
  \mnras, 449, 2

\bibitem[{{Meiring} {et~al.}(2013){Meiring}, {Tripp}, {Werk}, {Howk},
  {Jenkins}, {Prochaska}, {Lehner}, \& {Sembach}}]{Meiring2013}
{Meiring}, J.~D., {Tripp}, T.~M., {Werk}, J.~K., {Howk}, J.~C., {Jenkins},
  E.~B., {Prochaska}, J.~X., {Lehner}, N., \& {Sembach}, K.~R. 2013, \apj, 767,
  49

\bibitem[{{Miniati} {et~al.}(1999){Miniati}, {Ryu}, {Ferrara}, \&
  {Jones}}]{Miniati1999}
{Miniati}, F., {Ryu}, D., {Ferrara}, A., \& {Jones}, T.~W. 1999, \apj, 510, 726

\bibitem[{{Murray} {et~al.}(2005){Murray}, {Quataert}, \&
  {Thompson}}]{Murray2005}
{Murray}, N., {Quataert}, E., \& {Thompson}, T.~A. 2005, \apj, 618, 569

\bibitem[{{Oppenheimer} {et~al.}(2010){Oppenheimer}, {Dav{\'e}}, {Kere{\v s}},
  {Fardal}, {Katz}, {Kollmeier}, \& {Weinberg}}]{Oppenheimer2010}
{Oppenheimer}, B.~D., {Dav{\'e}}, R., {Kere{\v s}}, D., {Fardal}, M., {Katz},
  N., {Kollmeier}, J.~A., \& {Weinberg}, D.~H. 2010, \mnras, 406, 2325

\bibitem[{{Orlando} {et~al.}(2008){Orlando}, {Bocchino}, {Reale}, {Peres}, \&
  {Pagano}}]{Orlando2008}
{Orlando}, S., {Bocchino}, F., {Reale}, F., {Peres}, G., \& {Pagano}, P. 2008,
  \apj, 678, 274

\bibitem[{{Orlando} {et~al.}(2005){Orlando}, {Peres}, {Reale}, {Bocchino},
  {Rosner}, {Plewa}, \& {Siegel}}]{Orlando2005}
{Orlando}, S., {Peres}, G., {Reale}, F., {Bocchino}, F., {Rosner}, R., {Plewa},
  T., \& {Siegel}, A. 2005, \aap, 444, 505

\bibitem[{{Pittard} {et~al.}(2009){Pittard}, {Falle}, {Hartquist}, \&
  {Dyson}}]{Pittard2009}
{Pittard}, J.~M., {Falle}, S.~A.~E.~G., {Hartquist}, T.~W., \& {Dyson}, J.~E.
  2009, \mnras, 394, 1351

\bibitem[{{Pittard} {et~al.}(2010){Pittard}, {Hartquist}, \&
  {Falle}}]{Pittard2010}
{Pittard}, J.~M., {Hartquist}, T.~W., \& {Falle}, S.~A.~E.~G. 2010, \mnras,
  405, 821

\bibitem[{{Pittard} \& {Parkin}(2016)}]{Pittard2016}
{Pittard}, J.~M. \& {Parkin}, E.~R. 2016, \mnras, 457, 4470

\bibitem[{{Poludnenko} {et~al.}(2002){Poludnenko}, {Frank}, \&
  {Blackman}}]{Poludnenko2002}
{Poludnenko}, A.~Y., {Frank}, A., \& {Blackman}, E.~G. 2002, \apj, 576, 832

\bibitem[{{Rubin} {et~al.}(2014){Rubin}, {Prochaska}, {Koo}, {Phillips},
  {Martin}, \& {Winstrom}}]{Rubin2014}
{Rubin}, K.~H.~R., {Prochaska}, J.~X., {Koo}, D.~C., {Phillips}, A.~C.,
  {Martin}, C.~L., \& {Winstrom}, L.~O. 2014, \apj, 794, 156

\bibitem[{{Scannapieco}(2017)}]{Scannapieco2017}
{Scannapieco}, E. 2017, \apj, 837, 28

\bibitem[{{Scannapieco} \& {Br{\"u}ggen}(2010)}]{Scannapieco2010}
{Scannapieco}, E. \& {Br{\"u}ggen}, M. 2010, \mnras, 405, 1634

\bibitem[{{Scannapieco} \& {Br{\"u}ggen}(2015)}]{Scannapieco2015}
---. 2015, \apj, 805, 158

\bibitem[{{Schiano} {et~al.}(1995){Schiano}, {Christiansen}, \&
  {Knerr}}]{Schiano1995}
{Schiano}, A.~V.~R., {Christiansen}, W.~A., \& {Knerr}, J.~M. 1995, \apj, 439,
  237

\bibitem[{{Schneider} \& {Robertson}(2017)}]{Schneider2017}
{Schneider}, E.~E. \& {Robertson}, B.~E. 2017, \apj, 834, 144

\bibitem[{{Shin} {et~al.}(2008){Shin}, {Stone}, \& {Snyder}}]{Shin2008}
{Shin}, M.-S., {Stone}, J.~M., \& {Snyder}, G.~F. 2008, \apj, 680, 336

\bibitem[{{Sparre} {et~al.}(2019){Sparre}, {Pfrommer}, \&
  {Vogelsberger}}]{Sparre2019}
{Sparre}, M., {Pfrommer}, C., \& {Vogelsberger}, M. 2019, \mnras, 482, 5401

\bibitem[{{Sturm} {et~al.}(2011){Sturm}, {Gonz{\'a}lez-Alfonso}, {Veilleux},
  {Fischer}, {Graci{\'a}-Carpio}, {Hailey-Dunsheath}, {Contursi}, {Poglitsch},
  {Sternberg}, {Davies}, {Genzel}, {Lutz}, {Tacconi}, {Verma}, {Maiolino}, \&
  {de Jong}}]{Sturm2011}
{Sturm}, E., {Gonz{\'a}lez-Alfonso}, E., {Veilleux}, S., {Fischer}, J.,
  {Graci{\'a}-Carpio}, J., {Hailey-Dunsheath}, S., {Contursi}, A., {Poglitsch},
  A., {Sternberg}, A., {Davies}, R., {Genzel}, R., {Lutz}, D., {Tacconi}, L.,
  {Verma}, A., {Maiolino}, R., \& {de Jong}, J.~A. 2011, \apjl, 733, L16

\bibitem[{{Sur} {et~al.}(2016){Sur}, {Scannapieco}, \& {Ostriker}}]{Sur2016}
{Sur}, S., {Scannapieco}, E., \& {Ostriker}, E.~C. 2016, \apj, 818, 28

\bibitem[{{Tremonti} {et~al.}(2004){Tremonti}, {Heckman}, {Kauffmann},
  {Brinchmann}, {Charlot}, {White}, {Seibert}, {Peng}, {Schlegel}, {Uomoto},
  {Fukugita}, \& {Brinkmann}}]{Tremonti2004}
{Tremonti}, C.~A., {Heckman}, T.~M., {Kauffmann}, G., {Brinchmann}, J.,
  {Charlot}, S., {White}, S.~D.~M., {Seibert}, M., {Peng}, E.~W., {Schlegel},
  D.~J., {Uomoto}, A., {Fukugita}, M., \& {Brinkmann}, J. 2004, \apj, 613, 898

\bibitem[{{van Loo} {et~al.}(2007){van Loo}, {Falle}, {Hartquist}, \&
  {Moore}}]{vanLoo2007}
{van Loo}, S., {Falle}, S.~A.~E.~G., {Hartquist}, T.~W., \& {Moore}, T.~J.~T.
  2007, \aap, 471, 213

\bibitem[{{Veilleux} {et~al.}(2005){Veilleux}, {Cecil}, \&
  {Bland-Hawthorn}}]{Veilleux2005}
{Veilleux}, S., {Cecil}, G., \& {Bland-Hawthorn}, J. 2005, \araa, 43, 769

\bibitem[{{Waagan} {et~al.}(2011){Waagan}, {Federrath}, \&
  {Klingenberg}}]{Waagan2011}
{Waagan}, K., {Federrath}, C., \& {Klingenberg}, C. 2011, Journal of
  Computational Physics, 230, 3331

\bibitem[{{Wiersma} {et~al.}(2009){Wiersma}, {Schaye}, \&
  {Smith}}]{Wiersma2009}
{Wiersma}, R.~P.~C., {Schaye}, J., \& {Smith}, B.~D. 2009, \mnras, 393, 99

\bibitem[{{Xu} \& {Stone}(1995)}]{Xu1995}
{Xu}, J. \& {Stone}, J.~M. 1995, \apj, 454, 172

\end{thebibliography}

\clearpage

\end{document}